%% file: Bedouins2021.tex
\documentclass[12pt]{article}

\RequirePackage[OT1]{fontenc}
\usepackage{natbib}
\RequirePackage[colorlinks,citecolor=blue,urlcolor=blue]{hyperref}
\usepackage[english]{babel}
\usepackage{amsthm}
\usepackage{amsmath}
\usepackage{amsfonts}
\usepackage{amssymb}
\usepackage{graphicx}
\usepackage{enumerate}
\usepackage{color}
\usepackage{array}
\usepackage{psfrag,epsf}
\usepackage{url} 
\usepackage[export]{adjustbox}
\usepackage{bm}
\usepackage{titling}
\usepackage{xcolor}
\usepackage{colortbl}

\input{macros.tex}


\newcommand{\ffrac}[2]{\ensuremath{\frac{\displaystyle #1}{\displaystyle #2}}}

\colorlet{shadecolor}{gray!40}
\begin{document}
\thispagestyle{empty}
\baselineskip=28pt
\vskip 5mm
\begin{center} {\Large{\bf A combined statistical and machine learning \\ approach for spatial prediction of extreme \\ \vspace{-2mm} wildfire frequencies and sizes}}
\end{center}

\baselineskip=12pt
\vskip 5mm

\begin{center}
\large
Daniela Cisneros$^1$, Yan Gong$^1$, Rishikesh Yadav$^1$, \\ Arnab Hazra$^{1*}$, and Rapha{\"e}l Huser$^1$
\end{center}

\footnotetext[1]{
\baselineskip=10pt Computer, Electrical and Mathematical Sciences and Engineering (CEMSE) Division, King Abdullah University of Science and Technology (KAUST), Thuwal 23955-6900, Saudi Arabia. \\ $^*$E-mail: arnab.hazra@kaust.edu.sa}

\baselineskip=17pt
\vskip 4mm
\centerline{\today}
\vskip 6mm

\begin{center}
{\large{\bf Abstract}}
\end{center}

Motivated by the Extreme Value Analysis 2021 (EVA  2021) data challenge we propose a method based on statistics and machine learning for the spatial prediction of extreme wildfire frequencies and sizes. This method is tailored to handle large datasets, including missing observations. Our approach relies on a four-stage high-dimensional bivariate sparse spatial model for zero-inflated data, which is developed using stochastic partial differential equations (SPDE). In Stage 1, the observations are categorized in zero/nonzero categories and are modeled using a two-layered hierarchical Bayesian sparse spatial model to estimate the probabilities of these two categories. In Stage 2, before modeling the positive observations using a spatially-varying coefficients, smoothed parameter surfaces are obtained from empirical estimates using fixed rank kriging. This approximate Bayesian method inference was employed to avoid the high computational burden of large spatial data modeling using spatially-varying coefficients. In Stage 3, the standardized log-transformed positive observations from the second stage are further modeled using a sparse bivariate spatial Gaussian process. The Gaussian distribution assumption for wildfire counts developed in the third stage, is computationally effective but erroneous. Thus in Stage 4, the predicted values are rectified using Random Forests. Posterior inference is drawn for Stages 1 and 3 using Markov chain Monte Carlo (MCMC) sampling. A cross-validation scheme is then created for the artificially generated gaps, and the EVA 2021 prediction scores of the proposed model are compared to those obtained using certain natural competitors.

\baselineskip=16pt

\par\vfill\noindent

\pagenumbering{arabic}
\baselineskip=24pt

\newpage



\section{Introduction}
Wildfires have become one of the important concerns in recent years because they cause air pollution, extinction of floral and faunal species, significant economic loss, irreparable damage to the environment and the atmosphere, and threats to people's lives. Wildfires occur because of multiple causes, such as human intervention \citep[e.g., agricultural activities, campfires, and smoking,][]{pyne1996introduction}, lightning, volcanic eruption, debris burning, sparks from rock falls, and spontaneous combustion \citep{scott2000pre}. Typically, wildfires are accelerated by favorable conditions such as high flammability, extreme weather-like conditions (e.g., a drought), and the presence of combustible materials (e.g., forest matter). Recent wildfires in the United States (US) have led to considerable economic losses and social stresses \citep{brown2021us}. Moreover, there is concern that climate change may increase the intensity, duration, and frequency of wildfires \citep{abatzoglou2016impact, wuebbles2017climate, brown2021us}. Wildfire prediction is an important component of wildfire management because it impacts resource distribution, mitigation of adverse effects, and recovery efforts, and it is thus of crucial importance to develop resilient statistical methods that can reliably predict extreme wildfire events over space and time.

From a probabilistic viewpoint, wildfire occurrences and sizes can be seen as the results of random spatiotemporal processes; thus, it is important to understand their spatiotemporal distributions and underlying risk factors associated with them. In particular, wildfires can be linked to their spatial coordinates, e.g., the location of the fire origin, or the center of a burnt area, their temporal instant, and other relevant covariates. Extreme wildfires are attracting considerable attention because they are increasingly hazardous and often considered a more severe threat to ecosystems. 

Multiple statistical approaches have been proposed in the literature to predict the counts and sizes of wildfires using univariate probability models \citep{cumming2001parametric, preisler2004probability, preisler2007statistical, preisler2013forest, pereira2019statistical}. \cite{xi2019statistical} in particular made multiple important contributions to modeling fire risk components over recent decades, describing some key yet often overlooked fire characteristics, and they highlighted various areas of recent research that may enhance fire risk assesment models. 
However the spatial/temporal dependence of wildfires in their modeling approach was disregarded. To develop effective disaster management policies, statistical modeling of this dependence is important. Certain papers proposed using the so-called $K$ and $L$ functions for treating fire occurrences as spatial point-pattern datasets, \citep{genton2006spatio, hering2009modeling, juan2012pinpointing}. \cite{serra2012spatio} and \cite{roger2015rinla} treated these point-pattern datasets as gridded spatial datasets, in which the counts of fire occurrences (CNT) in each grid cell were modeled as spatially dependent count data; furthermore, the spatial dependence was modeled in terms of  a Gaussian process (GP), which is the most common tool for modeling spatial dependence because of its attractive theoretical and computational properties \citep{gelfand2016spatial}.  A log-Gaussian Cox process  \citep[LGCP,][]{moller1998log}, is a doubly stochastic construction, which compromises a Poisson point process at the data level, characterized with a random log-intensity modeled using a GP at the latent level. \cite{diggle2013spatial} reviewed the available literature on LGCPs in which LGCP models in different applications of spatiotemporal point pattern analysis were reported. Many researchers used LGCPs for occurrence modeling in a wide range of contexts; see, for example, \cite{serra2012spatio}, \cite{moller2010structured}, and \cite{gabriel2017detecting}. Furthermore, \cite{abdelfatah2016analytical} and \cite{trucchia2018surrogate} used GPs for modeling burnt area.

For large spatial datasets on a discretized spatial domain, continuos-space GPs are often replaced by a discrete Gaussian Markov random fields \citep[GMRFs,][]{rue2005gaussian}. GMRFs allow using sparse precision matrices, affording faster computations. Many researchers used joint analyses to study both fire occurrences and sizes in large datasets. \cite{rios2018studying} used a zero-inflated beta distribution, where a zero inflation was used to model the absence of fires, while the fraction of the burnt area was modeled using a beta distribution. Furthermore, the model parameters were assumed to be spatially varying, and GMRF priors were used in a fully Bayesian analysis. \cite{joseph2019spatiotemporal} compared certain probability distributions used for modeling frequencies and sizes of large wildfires to generate a posterior predictive distribution based on finite sample maxima for extreme events. The best performance was achieved using a zero-inflated negative binomial model for CNT and a lognormal model for burnt areas (BAs). Similarly, a marked LGCP model was proposed by \cite{pimont2021prediction} in which the authors modeled the occurrences using a point process and treated the fire sizes as marks. However, their approach estimated the model components for occurrences and fire sizes separately, which clearly limits the interaction between these two components. More recently, \cite{koh2021spatiotemporal} rectified this method by allowing joint estimation for all components using fully Bayesian inference.

Motivated by the 2021 Extreme Value Analysis conference (EVA 2021) data challenge, in which we participated as the team named \textit{The Bedouins}, we here propose an alternative approach based on statistics and machine learning (ML) for the spatial prediction of wildfire sizes or BAs and CNT at masked spatiotemporal locations. The complete dataset includes monthly observations (from March to September) at 3503 grid cells across the US between 1993 and 2015, in which each of the variables (i.e., BA and CNT) are masked at 80,000 spatiotemporal points. Because the spatial dimension is large and the dataset includes many zeros, a model that allows a scalable inferential scheme with high spatial dimension and zero inflation is required. Here, a four-stage high-dimensional bivariate spatial model is proposed for zero-inflated data. Our model is developed using stochastic partial differential equations (SPDEs). In Stage 1, the observations are categorized in zero/nonzero categories (zero BA indicates zero CNT and \textit{vice versa}). Moreover, a two-layered hierarchical Bayesian model is fitted, whereby the first layer is used for defining the zero/nonzero data in two categories of a real-valued latent process, and the second layer is used for developing an SPDE-based construction of the latent process. This model is used to estimate the probabilities of the categories at unobserved spatiotemporal sites. In Stage 2, before modeling the positive observations using a log-Gaussian process with spatially-varying parameters, the model parameters are empirically estimated (i.e., sample means and standard deviations (SDs) of the log-transformed positive observations) at each spatial location and smoothed parameter surfaces are obtained using fixed rank kriging \citep[FRK,][]{cressie2008fixed}. Fully Bayesian inference with spatially--varying parameters involves a large computational burden, which can be avoided using an efficient approximate Bayesian inference technique. In Stage 3, the standardized log-transformed positive observations from the second stage are modeled using a bivariate spatial GMRF. Despite the computational advantages of GP-based modeling, the model in Stage 3 erroneously assumes that the marginal distribution of CNT is Gaussian. Thus, the predicted values of CNT are finally corrected in Stage 4 using Random Forests (RF) in which BAs are treated as a covariate, and the missing BA values are imputed by the predicted BA values in Stage 3. Recently, ML algorithms, such as RF and neural networks (NN), have been successfully used to model fire occurrence data \citep{jain2020review}. In this study, CNT data are independently calibrated at each spatial location. Similarly, in \cite{saha2021random} was recently proposed a RF technique for spatially dependent data. Here, we use the simulation-based Markov Chain Monte Carlo (MCMC) method to draw posterior inference in Stages 1 and 3. This method involves moderate computational time even for high spatial dimension because of the sparse spatial structure implied by the SPDE. Our modeling framework targets wildfire prediction in the US; however, it is worth noting that it can be also adapted to other data scenarios with large spatial dimensions and zero inflation.

The paper is structured as follows: In Section \ref{sec:ExpAna}, an exploratory analysis of the US wildfire dataset is discussed. Section \ref{sec:Method} shows the development of an approach combining statistics and RF for the joint modeling of BA and CNT. In Section \ref{sec:Computation}, a brief overview of computational details is obtained. The proposed approach is applied to the US wildfire dataset and the results are then discussed in Section \ref{sec:DataAppl}. The conclusions of this study and perspectives for future research are presented in Section \ref{sec:Conclusions}.

\section{The US Wildfire dataset and exploratory analysis}
\label{sec:ExpAna}
In this section, the primary features of the US wildfire dataset are described and an exploratory graphical support for our modeling choices is provided.

\subsection{Data description and the missing data pattern} 
\label{subsec:data_description}

The wildfire dataset for the EVA 2021 data challenge comprises monthly observations at 3503 grid cells across the US Mainland using a spatial resolution of $0.5^\circ \times 0.5^\circ$. The original dataset (before masking) contains wildfire counts (CNTs) and aggregated burnt areas (BAs) in each pixel over a period of 23 years between (1993--2015) and for seven months per year (March to September). Information about 18 spatiotemporal land cover covariates (e.g., proportion of urban area, shrubland, grassland), 10 spatiotemporal meteorological covariates (e.g., temperature at 2 m above the ground, precipitation, evaporation of water), as well as certain purely spatial and temporal covariates (mean and standard deviation of the altitude, proportion of a pixel that is within the US Mainland, longitude and latitude for the center of the pixel, year, and month) was available. The organizers used the Shuttle Radar Topography Mission (SRTM) database available at a 90--m spatial resolution to calculate altitude-related covariates. A detailed description of the dataset is available at \citep{Opitz2022editorial}. 
The primary aim of this study was to estimate the predictive distribution function of BA and CNT at 28  severity thresholds as follows:
\begin{eqnarray} \label{severity_thresholds}
\nonumber \mathcal{U}_{\textrm{CNT}} &=& \{0,1,2,\ldots,9,10,12,14,\cdots,30,40,50,\cdots,100\},\\
\nonumber \mathcal{U}_{BA} &=& \{0,1,10,20,30,\ldots,100,150,200,250,300,400,500,1000, \\
&& 1500, 2000, 5000, 10000, 20000, 30000, 40000, 50000, 100000\}.
\end{eqnarray}
The final evaluation of all the registered teams was based on the weighted sum of the squared error between the empirical and predictive distribution functions, with weights given by $\tilde{\omega}_{\textrm{BA}}(u) = \tilde{\omega}_{\textrm{CNT}}(u) = 1 - (1 + (u + 1)^2 /1000)^{-1/4}$, $u\in\mathcal{U}_{\text{CNT}}$ or $\mathcal{U}_{\text{BA}}$, then rescaled to add up to 1.

\begin{figure}
    \centering
\adjincludegraphics[width = \linewidth, trim = {{.0\width} {.01\width} {.03\width} {.01\width}}, clip]{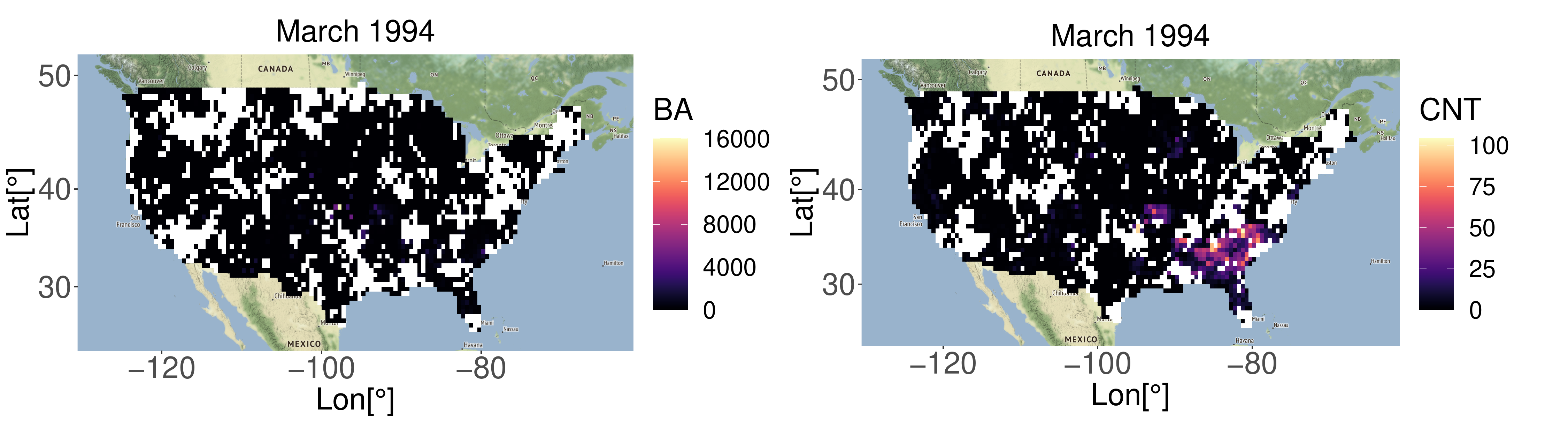}
    \caption{Spatial maps of BA (left) and CNT (right) in March 1994. The locations in which data are masked are highlighted in white.}
    \label{fig:plotNA}
\end{figure}

The original dataset does not contain any missing values. However, the organizers masked a total of 80,000 observations (14.18\%) across all the seven months of the alternative even years (1994, 1996, 1998, 2000, 2002, 2004, 2006, 2008, 2010, 2012, and 2014) to compare the spatial prediction performances of models proposed by participating teams. The masked spatiotemporal locations are not similar for BA and CNT; furthermore, the data were removed from low to high fire-prone regions. This demonstrates the requirement for modeling spatial dependence to borrow information from nearby pixels. Figure \ref{fig:plotNA} shows the spatial maps of BA and CNT in March 1994, where the masked data locations are highlighted in white. The masked observations were from small and large clusters of pixels and from regions in the southeast of the US, where high CNT values are observed close to masked pixels. Table \ref{table:zero} shows the proportions of masked observations of BA and CNT. As reported, either BA or CNT data or both are missing in 14.18\% of total cases. They are jointly missing in 8.68\% cases. Thus, BA and CNT information can be borrowed for the rest of 5.50\% non-missing observations by joint modeling, which leads to smaller standard errors of model parameters.

\begin{table}[h]
\begin{center}
\caption{The proportions of zero, nonzero, and missing values of BA and CNT in the total 563,983 observations.}\label{table:zero}
\begin{tabular}{|l|r|r|r|r|}
\hline
CNT \textbackslash BA & Zero  & Nonzero  & Missing & Total\\
\hline
Zero  & 49.60\% & 0.00\% & 3.34\% & 52.94\%\\
\hline
Nonzero  & 0.00\% & 30.70\% & 2.17\% & 32.87\%\\
\hline
Missing & 3.32\% & 2.18\% & 8.68\% & 14.18\%\\
\hline
Total & 52.93\% & 32.89\% & 14.18\% & 100.00\%\\
\hline
\end{tabular}
\end{center}
\end{table}

\subsection{Zero inflation} 

The high proportion of zero observations in the US wildfire dataset is one of its important characteristics; BA being zero/nonzero at a spatiotemporal location is equivalent to CNT being zero/nonzero. Table \ref{table:zero} lists the proportions of zero and nonzero values for BA and CNT. Using this equivalence, some of the missing data can be retrieved. CNT is zero in 3.34\% (out of the 14.18\%) of the cases where BA has missing value; hence, the BA values are zeros in these instances. Similarly, BAs are zero in 3.32\% (out of the 14.18\%) of the cases where CNT has missing values; hence, CNT values in these cases are zero. Thus, by filling these masked locations with zeros, a dataset with a smaller number of missing values can be obtained (specifically, 48,947 cases instead of 80,000 cases). Here, out of the final available observations, 61.61\% of the values are zeros. Therefore, the available information can be divided into two parts. In the first part, a spatiotemporal dataset of binary observations indicating whether BA/CNT is zero or not, is obtained; in the second part, only the positive values are maintained (treating zeros as missing data). For modeling positive BA and CNT, the logarithmic transformation can be used, and we call the resulting transformed dataset $\log$-BA and $\log$-CNT, respectively.

We now explore the covariate effects on the zero/nonzero indicators. The results of a probit regression model demonstrated significance for certain examined covariates at a significance level of 0.01. However, assuming a single regression coefficient for the entire spatiotemporal domain is not realistic. Despite increasing the computational burden, a cross-validation study does not demonstrate any significant improvement in the prediction performance after incorporating the available covariate information. Thus, to simplify calculations we choose to ignore the covariate information. Similarly, the spatiotemporal covariates can be ignored in the modeling of log-BA and log-CNT, unless they do not increase significantly the computational burden.

\begin{figure}
    \centering
\includegraphics[width=1\linewidth]{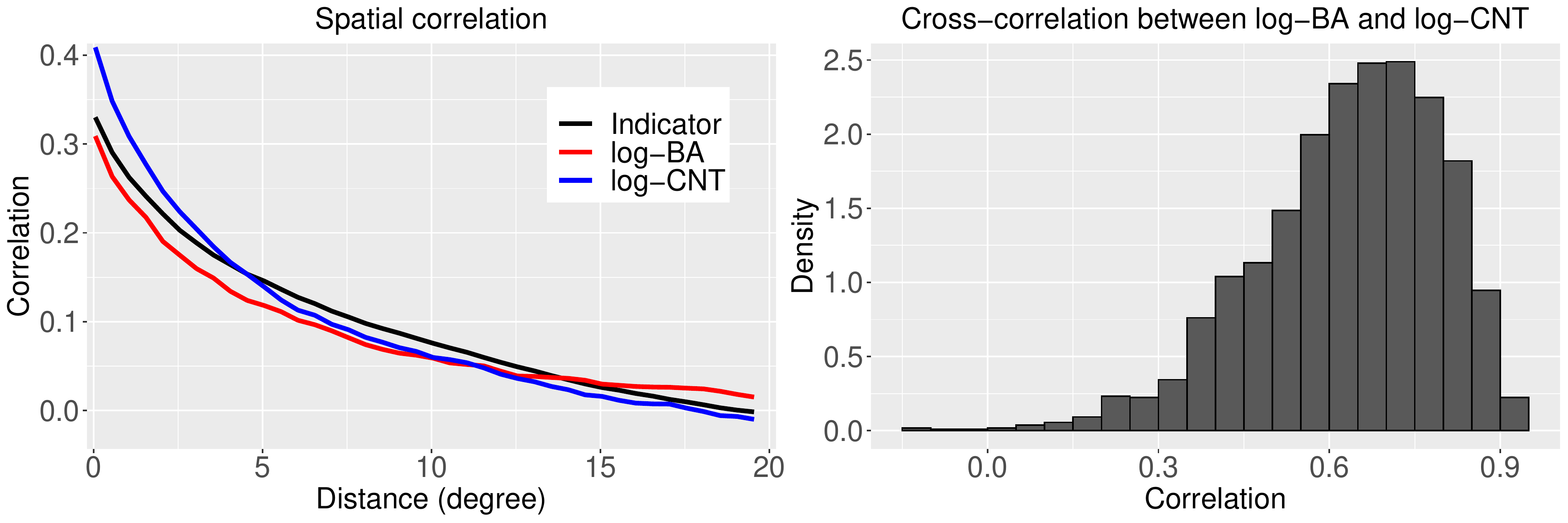}
 \caption{Left: empirical spatial correlation (smoothed) profiles of the zero/nonzero indicators (black), log-BA (red), and log-CNT (blue). Right: histogram of the empirical cross-correlation between log-BA and log-CNT across space.}
  \label{fig:Correlation}
\end{figure}

We then explore the spatial and temporal correlation profiles of the zero/nonzero indicators, log-BA, and log-CNT, as well as the cross-correlation between log-BA and log-CNT. The left panel of Figure \ref{fig:Correlation} shows (smoothed) empirical spatial correlation profiles as a function of distance. All three spatial correlation profiles decrease towards zero with the increase in geographical distance between pixels, and they demonstrate similar spatial range and small-scale variability (nugget effect). Thus, a reasonable separable correlation structure can be assumed for multivariate spatial modeling of log-BA and log-CNT. For every spatial location, the Lag-1 temporal autocorrelation was empirically calculated and was reported to be not significant at a significance level 0.01 for a large proportion of spatial locations (64.13\% locations for the zero/nonzero indicators, 96.39\% locations for log-BA, 89.46\% locations for log-CNT). A cross-validation study does not demonstrate any significant improvement in the prediction performance by incorporating temporal dependence. Thus, ignoring the temporal autocorrelation for all three spatiotemporal processes is reasonable, particularly considering the additional computational burden in a high spatial dimension. The cross-correlation between log-BA and log-CNT is empirically calculated at every spatial location (based on the temporal replicates), and the right panel of Figure \ref{fig:Correlation} shows the histogram of empirical cross-correlation values. The cross-correlation is quite high for most spatial locations, which indicates the requirement for joint statistical modeling of log-BA and log-CNT.

\begin{figure}
    \centering
    \includegraphics[width=1\linewidth]{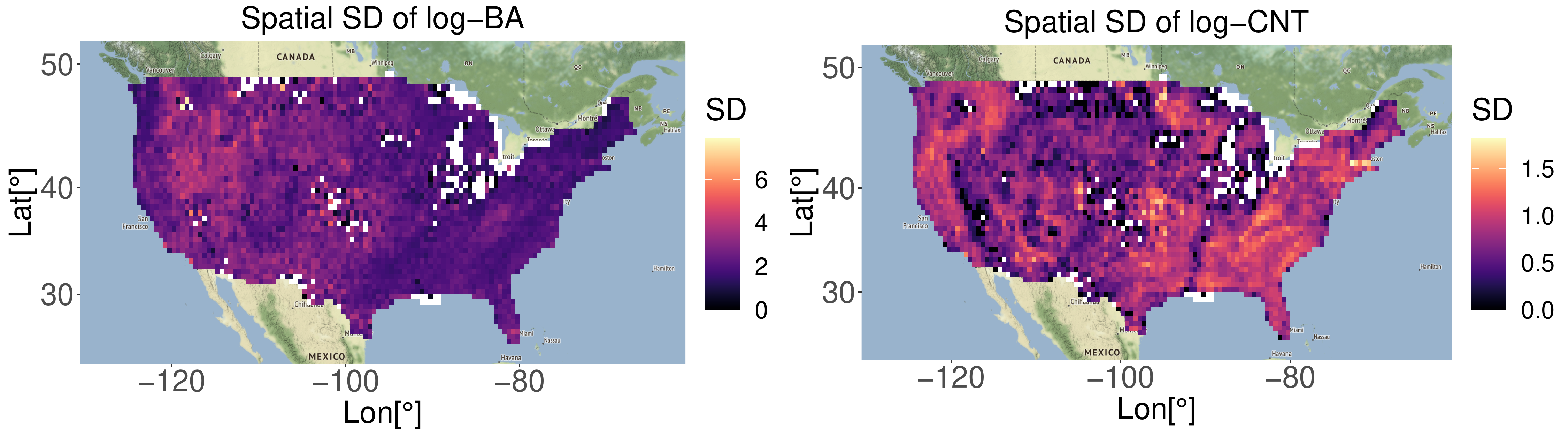}
    \caption{Pixel-wise standard deviation of log-BA (left) and log-CNT (right). Pixels where all observations are zero or that have only one positive observation (i.e. standard deviations are not available) are shown in white.}
    \label{fig:plotPositive}
\end{figure}

Finally, we explore the requirement for spatially-varying marginal distribution parameters for log-BA and log-CNT. When
log-BA and log-CNT are modeled using a probability distribution from a location-scale family, thus allowing a spatially-varying location profile is more common than allowing a spatially-varying scale profile, because of the computational benefits of the former. Figure \ref{fig:plotPositive} shows the empirical location-wise standard deviations of log-BA and log-CNT. For log-BA, standard deviations are generally lower in Eastern US than in Western US. For log-CNT, standard deviations are lower in the middle regions of US compared to the states closer to the Atlantic or the Pacific coasts. A similar spatial pattern is observed for the location-wise mean values of log-BA and log-CNT. Thus, a joint analysis of log-BA and log-CNT using a model with spatially-varying location and scale parameters is required.

\section{Joint modeling of burnt area (BA) and counts of fire occurrences (CNT)}
\label{sec:Method}

In this section, a four-stage model based on statistics and machine learning for the joint analysis of BA and CNT is described. As reported in Section \ref{sec:ExpAna}, the available information is divided into two parts. In the first part, a spatiotemporal dataset of binary observations is obtained to determine whether BA/CNT values are zero; in the second part, only the positive values are maintained, and log-BA and log-CNT are modeled on the logarithmic scale. In Stage 1, a sparse latent GP model for binary spatial data is proposed. For the other three stages, the modeling of log-BA and log-CNT using a combination of an approximate Bayesian inference technique and Random Forests (RFs) is described.

\subsection{Stage 1: A sparse latent Gaussian process model for wildfire occurrence data}
\label{subsec:Stage1}

The wildfire occurrence data are assumed independent and identically distributed (IID) across months and years; furthermore the observations across the US mainland for each month are assumed to only be spatially dependent. Thus,  by ignoring the month--year combinations, a generic notation is used for temporal replications.

For a spatial location $\bm{s}_i$ and time $t$, the BA and CNT values are denoted $\textrm{BA}_t(\bm{s}_i)$ and $\textrm{CNT}_t(\bm{s}_i)$, respectively, where $i \in \{1, \ldots, N\}$ and $t \in \{1, \ldots, T \}$, $N$ is the total number of pixels ($N=3503$), and $T$ is the total number of months ($T=161$). The wildfire occurrence indicator $Z_t(\bm{s}_i)$ at location $\bm{s}_i$ and time $t$ is defined as
\begin{align*}
   Z_t(\bm{s}_i) = \begin{cases}
   1, \hspace{.4cm}\textrm{if}\begin{cases}
  \parbox[t]{.8\textwidth}{$\textrm{BA}_t(\bm{s}_i)>0, ~\textrm{CNT}_t(\bm{s}_i) > 0$,}\\
    \text{$\textrm{BA}_t(\bm{s}_i)>0, \textrm{CNT}_t(\bm{s}_i)$ is missing, $~$\textrm{or}$~~$}\\
   \text{$\textrm{CNT}_t(\bm{s}_i) > 0, \textrm{BA}_t(\bm{s}_i)$ is missing,}
   \end{cases}\\ \vspace{-5mm}  \\
   0, \hspace{.4cm}\textrm{if}\begin{cases}
      \parbox[t]{.8\textwidth}{$\textrm{BA}_t(\bm{s}_i)=0, ~\textrm{CNT}_t(\bm{s}_i) = 0$,}\\
       \text{$\textrm{BA}_t(\bm{s}_i)=0, \textrm{CNT}_t(\bm{s}_i)$ is missing, $~$\textrm{or}$~~$}\\
   \text{$\textrm{CNT}_t(\bm{s}_i)=0, \textrm{BA}_t(\bm{s}_i)$ is missing,}
      \end{cases}\\
    $\texttt{NA},$~~~ \text{if ~~~$\textrm{BA}_t(\bm{s}_i)$ and $\textrm{CNT}_t(\bm{s}_i)$ are both missing.}   
    \end{cases}
\end{align*}


By assumption the replicated indicator processes, $Z_t(\cdot), t=1, \ldots, T$, are IID across time $t$, and we model $Z_t(\cdot)$ as 
\begin{eqnarray} \label{z_def}
    Z_t(\bm{s}_i) = \begin{cases}
        1, & \parbox[t]{.2\textwidth}{if~~$X_t(\bm{s}_i) > 0$}\\
        0, & \text{if~~$X_t(\bm{s}_i) < 0$},
    \end{cases}
    \textrm{where}~ X_t(\bm{s}_i) = \mu_Z(\bm{s}_i) + \varepsilon_t(\bm{s}_i),
\end{eqnarray}

and $\varepsilon_t(\cdot), t=1, \ldots, T$, are IID spatial GPs, in which $\textrm{E}[\varepsilon_t(\bm{s}_i)] = 0$ and $\textrm{Var}[\varepsilon_t(\bm{s}_i)] = 1$ for all $i \in \{1, \ldots, N\}$. The process $\varepsilon_t(\cdot)$ is assumed to follow an isotropic Mat\'ern spatial correlation (with nugget effect) given by
\begin{eqnarray} 
\label{cov_structure}
 \rho_{\varepsilon}(\bm{s}_i, \bm{s}_j) = \frac{r_{\varepsilon}}{\Gamma(\nu) 2^{\nu - 1}} \left( \frac{ d(\bm{s}_i, \bm{s}_j) }{\phi_{\varepsilon}} \right)^{\nu} K_{\nu} \left( \frac{ d(\bm{s}_i, \bm{s}_j) }{\phi_{\varepsilon}} \right) + (1 - r_{\varepsilon}) \mathbb{I}(\bm{s}_i = \bm{s}_j),
\end{eqnarray}

where $d(\bm{s}_i, \bm{s}_j)$ is the Euclidean distance between $\bm{s}_i$ and $\bm{s}_j$, $\phi_{\varepsilon} > 0$, $\nu > 0$ and $r_{\varepsilon} \in [0, 1]$ are the range, smoothness, and ratio of the spatial to total variation, respectively. In (\ref{cov_structure}), $K_{\nu}$ is the modified Bessel function of the degree $\nu$, and $\mathbb{I}(\bm{s}_i = \bm{s}_j) = 1$ if $\bm{s}_i = \bm{s}_j$, and 0 otherwise. When $r_{\varepsilon} = 1$, $\epsilon_t(\cdot)$ is mean-square differentiable if $\nu$ is an integer. For practical applications, identifying $\nu$ is difficult, and thus, it is generally fixed a priori. Here, the process $\varepsilon_t(\cdot)$ is not observable and hence estimating $\nu$ is more challenging. Therefore, we here set $\nu$ to one. To overcome the high computational burden due to the large spatial dimension, $\varepsilon_t(\cdot)$ is defined as a Gaussian Markov random field (GMRF) that has an (approximately) equivalent spatial covariance structure to the dense GP with a spatial Matérn $\rho_{\varepsilon}(\cdot, \cdot)$. As described in \cite{lindgren2011explicit}, the equivalence is derived from the one-to-one link between dense isotropic Matérn GPs and GMRFs. We next briefly summarize this link.

Suppose that $\varepsilon(\cdot)$ is a dense GP with correlation structure (\ref{cov_structure}) and $r_{\varepsilon}=1$. Then, $\varepsilon(\cdot)$ is the solution to the SPDE $(8 \phi_{\varepsilon}^{-2} - \Delta) \varepsilon(\bm{s}) = \mathcal{W}(\bm{s})$, where $\Delta = \frac{\delta^2}{\delta^2x} + \frac{\delta^2}{\delta^2y}$ is the Laplacian operator, $(8 \phi_{\varepsilon}^{-2} - \Delta)$ is a pseudo-differential operator, and $\mathcal{W}(\bm{s}) \overset{\textrm{IID}}{\sim} \textrm{Normal}(0,1)$. We can solve this SPDE using ﬁnite element methods \citep{ciarlet2002finite} over a triangular mesh in $\mathbb{R}^2$, where the triangles are formed following a Delaunay triangulation. Let the set of mesh nodes be denoted by $\mathcal{S}^* = \{ \bm{s}^*_1, \ldots, \bm{s}^*_{N^*} \}$. We construct a ﬁnite element representation of the solution to $\varepsilon(\bm{s}) = \sum_{j=1}^{N^*} \psi_j(\bm{s}) \varepsilon^*_j$ for some chosen basis functions $\psi_j(\cdot)$ and normally distributed weights $\varepsilon^*_j$ defined at the mesh nodes $\mathcal{S}^*$. We calculate the inner products $\langle\, \psi_j(\cdot), 1 \rangle$ and $\langle\, \nabla \psi_{j_1}(\cdot), \nabla \psi_{j_2}(\cdot) \rangle$, where $\langle\, f, g \rangle = \int f(\bm{s}) g(\bm{s}) d\bm{s}$, and obtain three $(N^* \times N^*)$-dimensional finite element matrices $\bm{C}$, $\bm{G}_1$, and $\bm{G}_2$. Here, $\bm{C}$ is a diagonal matrix, of which the $(j,j)^{th}$ entry $C_{j,j} = \langle\, \psi_j(\cdot), 1 \rangle$, $\bm{G}_1$ is a sparse matrix, of which the $(j_1,j_2)^{th}$ entry $G_{1_{j_1,j_2}} = \langle\, \nabla \psi_{j_1}(\cdot), \nabla \psi_{j_2}(\cdot) \rangle$, and $\bm{G}_2 = \bm{G}_1 \bm{C}^{-1} \bm{G}_1$. Further theoretical details are discussed in \cite{bakka2018spatial}. The vector $\bm{\varepsilon}^* = [\varepsilon^*_1, \ldots, \varepsilon^*_{N^*}]' \sim \textrm{Normal}_{N^*}(\bm{0}, \bm{Q}_{\phi_{\varepsilon}}^{-1})$, where the precision matrix is $\bm{Q}_{\phi_{\varepsilon}} = (4\pi)^{-1} \phi_{\varepsilon}^2 \left[\phi_{\varepsilon}^{-4} \bm{C} + 2 \phi_{\varepsilon}^{-2} \bm{G}_1 + \bm{G}_2 \right]$. To project $\bm{\varepsilon}^*$ back to the data locations $\mathcal{S}$, we evaluate $a_{ij} = \psi_j(\bm{s}_i)$ for each location $\bm{s}_i$ and mesh node $\bm{s}_j^*$. The $(N\times N^*)$-dimensional matrix $\bm{A}$, of which $(i,j)^{th}$ entry is $a_{ij}$, is called the SPDE projection matrix, and $\bm{A} \bm{\varepsilon}^* \sim \textrm{Normal}_{N}(\bm{0}, \bm{A}\bm{Q}_{\phi_{\varepsilon}}^{-1}\bm{A}')$. The covariance matrix $\bm{A}\bm{Q}_{\phi_{\varepsilon}}^{-1}\bm{A}'$ approximates the Mat\'ern correlation matrix obtained by evaluating (\ref{cov_structure}) at $\mathcal{S}$ (for $r_{\varepsilon}=1$).

\begin{figure}[h]
    \centering
    \begin{minipage}{0.49\textwidth}
        \centering
\includegraphics[width=1\linewidth]{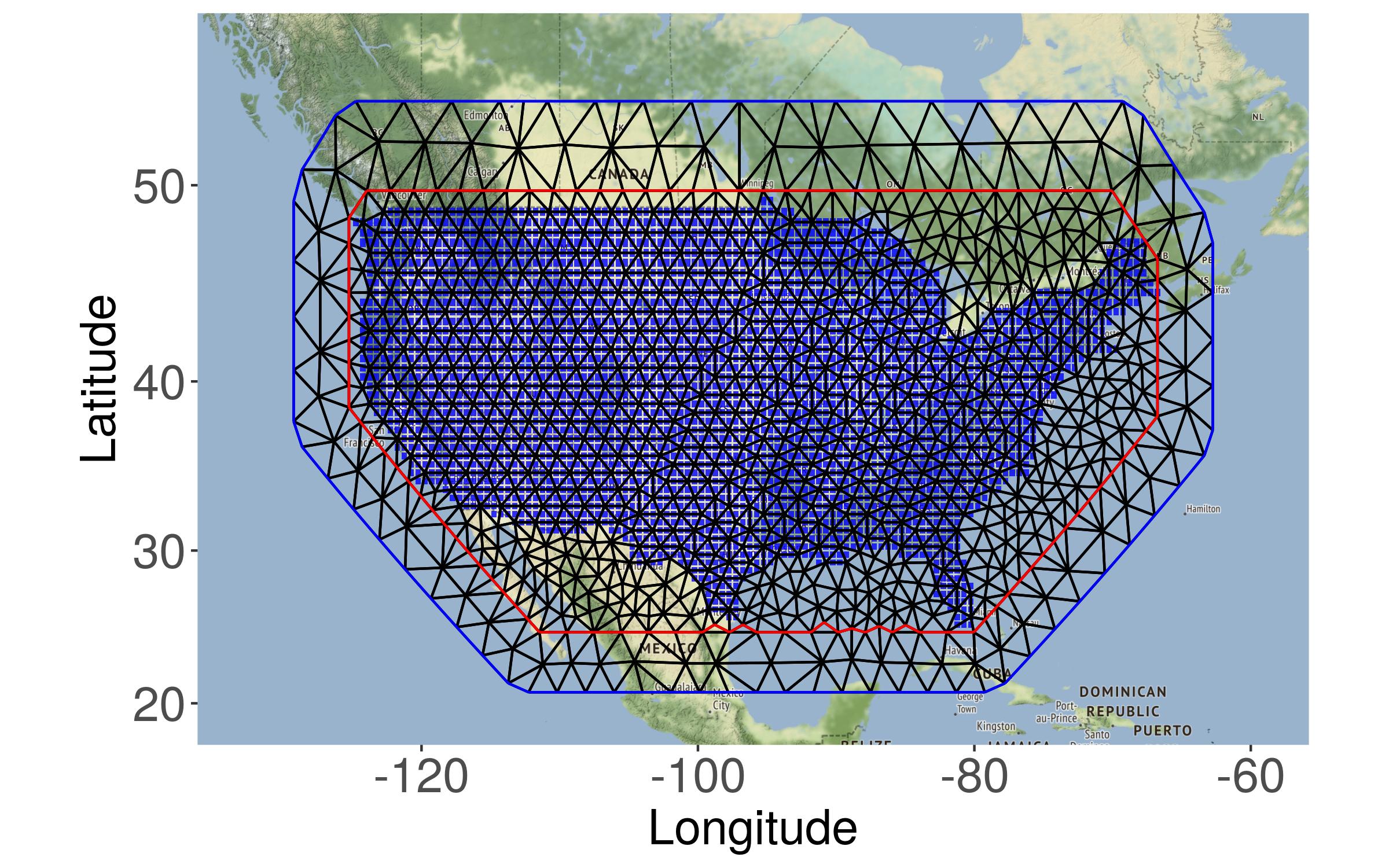}
    \end{minipage}\hfill
    \begin{minipage}{0.49\textwidth}
        \centering
\includegraphics[width=1\linewidth]{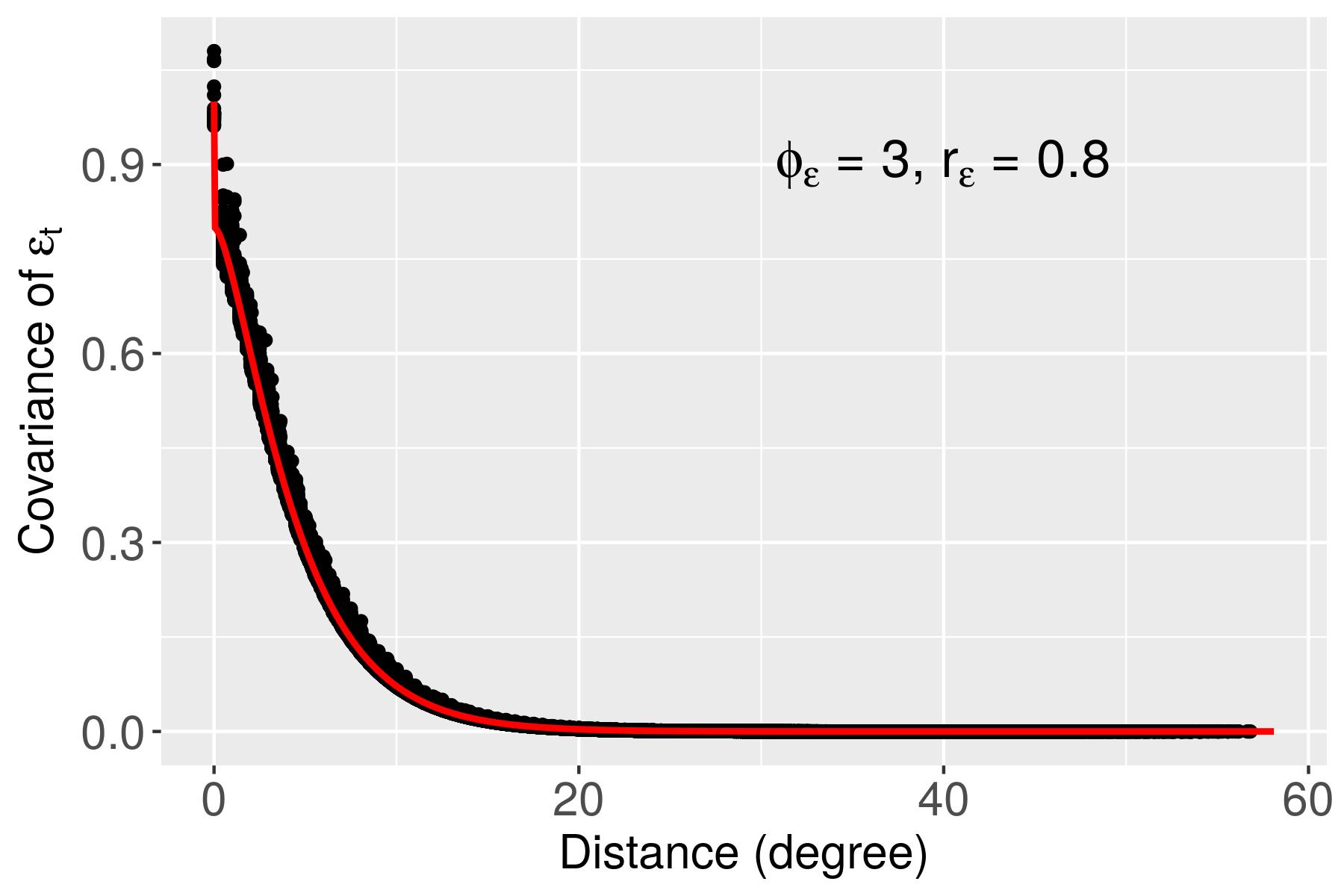}        
    \end{minipage}
    \caption{Left: triangulated mesh over the United States, which is used to develope the spatial process $\varepsilon_t(\cdot)$, using SPDEs. Right: comparison of a possible Mat\'ern correlation structure (red) and the pairwise covariances (as a function of distance) between two spatial locations obtained from the SPDE approximation (black dots).}
    \label{fig:Mesh_Var}
\end{figure}

Suppose that for each $t\in \{ 1, \ldots, T\}$, $\bm{\varepsilon}^*_t$ is an independent copy of $\bm{\varepsilon}^*$. We construct a vector $\bm{\varepsilon}_t = [\varepsilon_t(\bm{s}_1), \ldots, \varepsilon_t(\bm{s}_N)]'$ from $\bm{\varepsilon}^*_t$ as
\begin{eqnarray} \label{z_construction}
\bm{\varepsilon}_t = \sqrt{r_{\varepsilon}} \bm{A} \bm{\varepsilon}^*_t + \sqrt{1-r_{\varepsilon}} \tilde{\bm{\varepsilon}}_t,
\end{eqnarray}
where $\tilde{\bm{\varepsilon}}_t = [\tilde{\varepsilon}_t(\bm{s}_1), \ldots, \tilde{\varepsilon}_t(\bm{s}_N)]'$ with $\tilde{\varepsilon}_t(\bm{s}_i) \overset{\textrm{IID}}{\sim} \textrm{Normal}(0, 1)$. The final covariance matrix of $\bm{\varepsilon}_t$ is $\bm{\Sigma}_{\varepsilon} = r_{\varepsilon} \bm{A} \bm{Q}_{\phi_{\varepsilon}}^{-1} \bm{A}' + (1-r_{\varepsilon})\bm{I}_N$, and it approximates the Mat\'ern correlation matrix obtained by evaluating (\ref{cov_structure}) at $\mathcal{S}$, for any $r_{\varepsilon} \in [0, 1]$. We then discuss the GMRF approximation accuracy in our data application, for the values of $\phi_{\varepsilon}$ and $r_{\varepsilon}$ set to 3 and 0.8 respectively. These
values are similar to estimates obtained in our data application. The left panel of Figure \ref{fig:Mesh_Var} shows the mesh that we used for the US wildfire data analysis; here, there are $N^* = 1027$ mesh nodes. In this SPDE mesh, we calculate the covariance between every pair of spatial locations (the elements of the matrix $\bm{\Sigma}_{\varepsilon}$) and the corresponding true Mat\'ern correlation based on (\ref{cov_structure}). The results of these calculations are presented as a function of distance in the right panel of Figure \ref{fig:Mesh_Var}. The true correlation structure of dense GP is well approximated by the corresponding GMRF. Moreover, the sparsity of $\bm{Q}_{\phi_{\varepsilon}}$ can be exploited to allow quick computations. Moreover, while categorizing the latent process given in (\ref{z_def}), the conditional spatial independence structure $\bm{\varepsilon}_t \vert \bm{\varepsilon}^*_t \sim \textrm{Normal}_{N}(\sqrt{r_{\varepsilon}} \bm{A} \bm{\varepsilon}^*_t, (1-r_{\varepsilon}) \bm{I}_N)$ allows the univariate imputation of latent variables, which is exploited for spatial prediction.

After marginalization with respect to $X_t(\bm{s}_i)$, we obtain that $\textrm{Pr}(Z_t(\bm{s}_i) = 1) \approx \Phi(\mu_Z(\bm{s}_i))$, where $\Phi(\cdot)$ is the standard normal distribution function. The conditional distribution of $Z_t(\bm{s}_i)$ given $\bm{\varepsilon}^*_t$ is $\textrm{Pr}(Z_t(\bm{s}_i) = 1 \vert \bm{\varepsilon}^*_t) = \Phi((\mu_Z(\bm{s}_i) + \sqrt{r_{\varepsilon}} \bm{a}'_i \bm{\varepsilon}^*_t) / \sqrt{1-r_{\varepsilon}})$, where $\bm{a}_i$ is the $i^{th}$ row of $\bm{A}$, and $\textrm{Pr}(Z_t(\bm{s}_i) = 0 \vert \bm{\varepsilon}^*_t) = 1 - \textrm{Pr}(Z_t(\bm{s}_i) = 1 \vert \bm{\varepsilon}^*_t)$. The variables $Z_t(\bm{s}_i)$ and $Z_t(\bm{s}_j)$ are conditionally independent given $\bm{\varepsilon}^*_t$. There is no closed form expression for the joint distribution of $Z_t(\bm{s}_i)$ and $Z_t(\bm{s}_j)$, after marginalizing with respect to $X_t(\bm{s}_i)$.
\subsection{Stage 2: Approximate Bayesian inference to smooth spatially-varying parameters}
\label{subsec:Stage2} 
Despite the discreteness of CNTs,  we model log-BA and log-CNT using a GP with spatially-varying location and scale terms, because of the computational attractiveness of GPs, as follows
\begin{eqnarray} \label{eq:bivarmodel}
\nonumber \log\text{-BA}_t(\bm{s}_i) &=& \mu_{1}(\bm{s}_i) + \sigma_1(\bm{s}_i) W_{t1}(\bm{s}_i), \\
\log\text{-CNT}_t(\bm{s}_i) &=& \mu_{2}(\bm{s}_i) + \sigma_2(\bm{s}_i) W_{t2}(\bm{s}_i), 
\end{eqnarray}
where  $\nonumber \bm{W}_{t}(\bm{s}_i) = [W_{t1}(\bm{s}_i), W_{t2}(\bm{s}_i)]'$ is a bivariate standard GP (zero mean and unit variance for all marginal distributions). A full Bayesian inference is computationally challenging, and thus, in Stage 2, we focus on estimating the parameter surfaces $\mu_{1}(\cdot)$, $\sigma_1(\cdot)$, $\mu_{2}(\cdot)$, and $\sigma_2(\cdot)$ only, while the parameters within the bivariate spatial GP $\bm{W}_t(\cdot)$ are treated as nuisance parameters. Here, the procedure for $\mu_{1}(\cdot)$, which is the same procedure used to estimate the surfaces $\log[\sigma_1(\cdot)]$, $\mu_{2}(\cdot)$, and $\log[\sigma_2(\cdot)]$ is detailed; all surfaces evaluated at any $\bm{s}_i$ are defined over the whole real line.

An approximate Bayesian inference scheme, similar to Max-and-Smooth  \citep{hrafnkelsson2021max, johannesson2019approximate}, is applied to obtain parameter surfaces in two steps. In the first step, we estimate $\mu_{1}(\bm{s}_i)$, $\sigma_1(\bm{s}_i)$, $\mu_{2}(\bm{s}_i)$, and $\sigma_2(\bm{s}_i)$ at each $\bm{s}_i$ separately, using location-wise empirical means and standard deviations, ignoring any spatial/temporal trend or dependence. We denote the estimates by $\hat{\mu}_{1}(\bm{s}_i)$, $\log[\hat{\sigma}_1(\bm{s}_i)]$, $\hat{\mu}_{2}(\bm{s}_i)$, and $\log[\hat{\sigma}_2(\bm{s}_i)]$.

In the second step we smooth parameter surfaces by treating the preliminary estimates as noisy measurements of the true underlying parameters. Specifically, for the parameter surface $\mu_{1}(\cdot)$, we assume that $\hat{\mu}_{1}(\bm{s}_i) = \mu_{1}(\bm{s}_i) + e(\bm{s}_i)$, where $\mu_{1}(\bm{s}_i)$ is the true parameter value; furthermore the estimate $\hat{\mu}_{1}(\bm{s}_i)$ is perturbed from the true parameter by a pure nugget term $e(\bm{s}_i) \sim \textrm{Normal}(0, \sigma^2_e)$. Because of the large spatial dimension, the prior for $\mu_{1}(\bm{s}_i)$ is assumed to follow a low-rank structure as
\begin{equation}
    \mu_{1}(\bm{s}_i) = \beta_0 + \beta_1 \textrm{lon}(\bm{s}_i) + \beta_2 \textrm{lat}(\bm{s}_i) + \sum_{r=1}^{3} \sum_{k=1}^{K_r} h_{rk}(\bm{s}_i) \omega^*_{rk} + \xi(\bm{s}_i), ~~\bm{s}_i \in \mathcal{S},
\end{equation}
where $h_{rk}(\cdot)$ is the $k^{th}$ spatial Gaussian kernel at the $r^{th}$ resolution, the spatial random effects are $\bm{\omega}^*_{r} = [\omega^*_{rk_1}, \ldots, \omega^*_{rk_r}]'\overset{\textrm{Indep}}{\sim} \textrm{Normal}_{K_r}(\bm{0}, \bm{\Sigma}(\bm{\theta}_r))$, and $\xi(\bm{s}) \overset{\textrm{IID}}{\sim} \textrm{Normal}(0, \sigma^2_{\xi})$. By selecting flat priors for hyperparameters, the estimation of $\mu_{1}(\bm{s}_i), \bm{s}_i \in \mathcal{S}$, lies in the setting of fixed rank kriging (FRK) using a frequentist approach, which is readily implemented using the \texttt{R} package \texttt{FRK} \citep{JSSv098i04}. For certain pixels, there was no available positive observation. In these instances $\hat{\mu}_{1}(\bm{s}_i)$ is treated as missing and the corresponding values of $\mu_{1}(\bm{s}_i)$ are predicted  based on the available first-step estimates. We repeat the same procedure for other parameter surfaces and obtain smoothed estimates, say, $\tilde{\mu}_{1}(\cdot)$, $\tilde{\sigma}_1(\cdot)$, $\tilde{\mu}_{2}(\cdot)$, and $\tilde{\sigma}_2(\cdot)$. Finally, we also obtain $\bm{W}_{t}(\bm{s}_i), \bm{s}_i \in \mathcal{S}, t=1, \ldots, T$, by plugging the smoothed estimates in (\ref{eq:bivarmodel}). 


\subsection{Stage 3: Bivariate spatial modeling of standardized log-BA and log-CNT}
\label{subsec:Stage3}
In this stage, we model the standardized variables $\widehat{W}_{t1}(\bm{s}_i)=\{ \log\text{-BA}_t(\bm{s}_i)-\mu_1(\bm{s}_i)\}/ \sigma_1(\bm{s}_i)$ and $\widehat{W}_{t2}(\bm{s}_i)=\{ \log\text{-CNT}_t(\bm{s}_i)-\mu_2(\bm{s}_i)\}/ \sigma_2(\bm{s}_i)$, obtained in Stage 2 using (\ref{eq:bivarmodel}). Suppose that $\widehat{\bm{W}}_{tp} = [\widehat{W}_{tp}(\bm{s}_1), \ldots, \widehat{W}_{tp}(\bm{s}_N)]'$ for $p=1,2$, and $\widehat{\bm{W}}_{t} = [\widehat{\bm{W}}'_{t1}, \widehat{\bm{W}}'_{t2}]'$. We model $\widehat{\bm{W}}_{t}$ as $\widehat{\bm{W}}_{t} = \left[ \bm{I}_2 \otimes \bm{A} \right] \bm{\eta}^*_{t} + \tilde{\bm{\eta}}_{t}$, where
\begin{eqnarray}
	\nonumber \bm{\eta}^*_{t} \sim \textrm{Normal}_{2N}\left( \bm{0}_{2N}, r_{\eta} \left(
	\begin{array}{c}
	1 ~~ \rho_{\eta} \\
	\rho_{\eta} ~~ 1
	\end{array}
	\right) \otimes \bm{Q}^{-1}_{\phi_\eta} \right),~
\end{eqnarray}

$$\tilde{\bm{\eta}}_{t} \sim \textrm{Normal}_{2N}\left( \bm{0}_{2N}, (1 - r_{\eta}) \left(
	\begin{array}{c}
	1 ~~ \rho_{\eta} \\
	\rho_{\eta} ~~ 1
	\end{array}
	\right) \otimes \bm{I}_{N} \right).$$

Similarly to Stage 1, we model the processes $\widehat{W}_{tp}(\cdot)$ using GMRFs. The same SPDE mesh and SPDE projection matrix as those used in Stage 1 are used. The spatial correlation of each component $\widehat{W}_{tp}(\cdot)$ is approximately equal to (\ref{cov_structure}), with $\phi_{\varepsilon}$ and $r_{\varepsilon}$ replaced by $\phi_{\eta}$ and $r_{\eta}$, respectively. This is is confirmed by similar empirical spatial correlation profiles to those of log-BA and log-CNT shown in Figure \ref{fig:Correlation}. The marginal standard deviation of each $\widehat{W}_{tp}(\bm{s}_i)$ is approximately equal to one (as described in Stage 1); thus, the cross-covariance between $\widehat{W}_{t1}(\bm{s}_i)$ and $\widehat{W}_{t2}(\bm{s}_i)$ is approximately equal to the cross-correlation $\rho_{\eta} \in [-1, 1]$ for each $\bm{s}_i$ and $t$. To summarize, an approximately separable bivariate GMRF, where the marginal distributions have zero mean and unit variance, is defined.

At a spatiotemporal prediction location $(\bm{s}_i, t)$, we need to simulate the missing process $\widehat{W}_{tp}(\bm{s}_i)$, possibly for both $p=1,2$. Let the elements of $2N$-length vector $\left[ \bm{I}_2 \otimes \bm{A} \right] \bm{\eta}^*_t$ be denoted by $[\bm{\tilde{\eta}}^{*'}_{t1}, \bm{\tilde{\eta}}^{*'}_{t2}]'$, where $\bm{\tilde{\eta}}^{*}_{t1} = [\tilde{\eta}^*_{t1}(\bm{s}_1), \ldots, \tilde{\eta}^*_{t1}(\bm{s}_N)]'$ and $\bm{\tilde{\eta}}^{*}_{t2} = [\tilde{\eta}^*_{t2}(\bm{s}_1), \ldots, \tilde{\eta}^*_{t2}(\bm{s}_N)]'$. Then, 
\begin{eqnarray}
	\widehat{\bm{W}}_t(\bm{s}_i) \sim \textrm{Normal}_{2}\left( \left(
	\begin{array}{c}
	\tilde{\eta}^*_{t1}(\bm{s}_i)  \\
	 \tilde{\eta}^*_{t2}(\bm{s}_i)
	\end{array}
	\right), (1 - r_{\eta}) \left(
	\begin{array}{c}
\nonumber 1 ~~ \rho_{\eta} \\
	\rho_{\eta} ~~ 1
	\end{array}
	\right) \right).
\end{eqnarray}

The predicted values are denoted by $\tilde{W}_{tp}(\bm{s}_i)$, possibly for both $p=1$ and $p=2$. Then, the predicted values of log-BA and log-CNT are obtained by plugging $\tilde{\mu}_{1}(\cdot)$, $\tilde{\sigma}_1(\cdot)$, $\tilde{\mu}_{2}(\cdot)$, and $\tilde{\sigma}_2(\cdot)$ (obtained in Stage 2) and $\tilde{W}_{tp}(\bm{s}_i)$ (obtained in Stage 3) into (\ref{eq:bivarmodel}). The distribution functions of BA and CNT at $(\bm{s}_i, t)$ are approximately $\textrm{Lognormal}(\mu_{1}(\bm{s}_i), \sigma^2_1(\bm{s}_i))$ and $\textrm{Lognormal}(\mu_{2}(\bm{s}_i), \sigma^2_2(\bm{s}_i))$ distribution functions, and they are denoted $F_{LN}(\cdot~;~ \mu_{1}(\bm{s}_i), \sigma^2_1(\bm{s}_i))$ and $F_{LN}(\cdot~;~ \mu_{2}(\bm{s}_i), \sigma^2_2(\bm{s}_i))$, respectively. Furthermore, by incorporating zero inflation, $F_{\textrm{BA}}$ and $F_{\textrm{CNT}}$, the distribution functions of BA and CNT at $(\bm{s}_i, t)$, respectively, are given by (up to the SPDE approximation),
\begin{eqnarray} \label{eq:bivarmodel_cdf}
\nonumber F_{\textrm{BA}}(x;~ \mu_{Z}(\bm{s}_i), \mu_{1}(\bm{s}_i), \sigma^2_1(\bm{s}_i)) &\approx& 1 - \Phi(\mu_Z(\bm{s}_i)) \\ \nonumber 
&+& \Phi(\mu_Z(\bm{s}_i))F_{LN}(x; \mu_{1}(\bm{s}_i), \sigma^2_1(\bm{s}_i)), \\
\nonumber F_{\textrm{CNT}}(x;~ \mu_{Z}(\bm{s}_i), \mu_{2}(\bm{s}_i), \sigma^2_2(\bm{s}_i)) &\approx& 1 - \Phi(\mu_Z(\bm{s}_i)) \\ \nonumber 
&+& \Phi(\mu_Z(\bm{s}_i)) F_{LN}(x;\mu_{2}(\bm{s}_i), \sigma^2_2(\bm{s}_i)), x \geq 0. 
\end{eqnarray}


\subsection{Stage 4: Random Forests (RFs)}
\label{subsec:Stage4}

Until Stage 3, only the statistical approaches for joint modeling of BA and CNT, were considered. In the joint modeling, the stage-specific models did not incorporate any covariate information (except in certain priors that are discussed later). As reported in Section \ref{sec:ExpAna}, incorporating covariates in the statistical modeling framework and allowing for spatially-varying regression coefficients is computationally challenging and do not significantly improve the prediction performance. Moreover, assuming that the marginal distributions of log-CNT are Gaussian is computationally beneficial; however, it is inaccurate because of the discrete nature of CNT. Thus, in Stage 4, a rectification of the predicted values of CNT obtained from Stages 1 through 3 is proposed. Typically, machine learning (ML) approaches (such as RFs) do not have any distributional assumptions and involve low computational burden, while allowing a straightforward incorporation of covariate information.

As reported, the prediction performance is not improved using a simple regression setting. Nevertheless, for interpretation and prediction purposes, it is important to incorporate meaningful covariates, such as some landcover types and climate/weather conditions, because they play a key role in the occurrence of wildfires, as shown in several studies \citep{fusco2019invasive, nadeem2020mesoscale}. These covariates are interrelated and show seasonal dependence, but classical statistical methods, including generalized linear models, cannot account for the sophistication in this process. \cite{jain2020review} reviewed extensive literature (until 2019) on ML application in wildfire science and management, in which they reported RFs to be a more common method for predicting fire occurrences since 2012, before models based on artificial neural networks (NNs) and support vector machines (SVMs) were widely reported in the literature. It is reasonable to use RFs in the rectification of the US wildfire data analysis, because tree-based models are often suitable for classification problems, hence they are suitable for the discrete data CNT, and RF generates nonlinear regression models, which facilitates correct formulations of covariates.

The RF algorithm \citep{breiman2001random} involves an ensemble of many decision trees, in which individual trees are trained based on a random subset of the data, and are drawn with replacement. A random subset (resampled with replacement)  of covariates is selected at each node of every decision tree. For classification problems, each individual tree forms a class; furthermore, the predicted class is selected by the highest votes. The success of RFs are attributed to the low correlation among trees (each tree is trained independently from others); thus, it results in low prediction variance. Importantly, RFs minimize the correlation between trees; hence it provides higher accuracy than compared to individual trees, which explains the success of classification and regression trees (CART).

For the EVA 2021 data challenge, the prediction performance is evaluated based on score functions with high weights to the data categorized into higher severity levels as described in Section \ref{sec:ExpAna}. Considering the size of the dataset and underlying computational cost, this specific task (i.e., that of the challenge) was converted in classification problem, in which CNT was classified into 29 categories, (based on the 28 severity thresholds provided for model evaluation in \ref{severity_thresholds}) and labeled at each spatiotemporal location. Then CNT is converted into categorical data, which inevitably decreased in resolution and essentially did not affect the ability to achieve high prediction accuracy. Moreover, BA is here included as a covariate, in which the missing values of BA are imputed by the joint statistical modeling of BA and CNT.

\subsection{Modeling based on log-Gaussian Cox processes (LGCPs): another competing approach for CNT}
\label{subsec:lgcp}

Log-Gaussian Cox processes (LGCPs),  or Cox process, are a commonly flexible approach for analyzing point pattern data and are obtained by assuming a hierarchical Bayesian structure. At the first level of this structure (data level), the response at every discretized location is assumed to follow a Poisson distribution conditioned on its random intensity measure; at the second level (process level), the intensity measure is assumed to follow a log-Gaussian process defined over the spatial domain. In particular, a spatial point process $Y(\cdot)$ defined over $\mathcal{S} \subset \mathbb{R}^2$ is called a Cox process if $Y(\bm{s}) \vert \Lambda(\bm s) \overset{\textrm{Indep}}{\sim} \textrm{Poisson}(\exp[\Lambda(\bm s)])$ and the log-intensity process $\{\Lambda(\bm s); \bm s\in \mathcal{S}\}$ is a GP.

To fit the LGCP model in a tractable manner, a common approach is to divide the whole spatial region in a uniform grid and assume a constant log-intensity in each grid cell. Here, the US wildfire dataset is already gridded, and $\{\bm{s}_i, i=1, \ldots, N\}$ denote the spatial locations (the centroids of the grid cells) as defined in Section \ref{subsec:Stage1}. The log-intensity vector of the $t$-th time point is denoted $\bm \Lambda_t=[\Lambda_t(\bm s_1),\ldots,\Lambda_t(\bm s_N)]'$. We have $\bm \Lambda_t\sim \textrm{Normal}_N(\bm X_{\Lambda}\bm\beta_{\Lambda}, \sigma_{\Lambda}^2 \bm C_{\Lambda})$, where $\bm X_{\Lambda}$ is the design matrix, $\bm \beta_{\Lambda}$ is the vector of regression coefficients, $\sigma_{\Lambda}^2$ is the common variance term, and $\bm C_{\Lambda}$ is the corresponding correlation matrix. For a dense correlation matrix $\bm C_{\Lambda}$, the estimation procedure involves a high computational burden and thus, similar to Section \ref{subsec:Stage1}, we use an SPDE approximation-based construction of $\bm C_{\Lambda}$. We denote the CNT data at spatial location $\bm{s}_i$ and time $t$ by $\textrm{CNT}_t(\bm{s}_i)$ and propose the following LGCP model: 
\begin{eqnarray}
\label{eq:LGCP}
\textrm{CNT}_t(\bm s_i) \mid \Lambda_t(\bm s_i) &\overset{\textrm{Indep}}{\sim}& \text{Poisson} \left(\exp[\Lambda_t(\bm s_i)]\right), \quad i=1,\ldots,N, \, t=1,\ldots T, \nonumber  \\
 \Lambda_t(\bm s_i) &=& \mu_t(\bm s_i) + \zeta_t(\bm s_i),
\end{eqnarray}
where $\zeta_t(\cdot), t = 1, \ldots,T$, are IID copies from a GMRF that has an (approximately) equivalent isotropic Mat\'ern spatial correlation given by \eqref{cov_structure} (see Section \ref{subsec:Stage1} for more details), with $\phi_{\varepsilon}$ and $r_{\varepsilon}$ replaced by  $\phi_{\zeta}$ and $r_{\zeta}$, respectively, and $\mu_t(\bm s_i)$ denotes the spatiotemporal mean process at spatial location $\bm{s}_i$ and time $t$ and is defined in terms of fixed covariates. Multiple combinations of covariates (i.e., all covariates, only spatial covariates, including/excluding BA) are compared and discussed in Section \ref{sec:DataAppl}.


\section{Computation}
\label{sec:Computation}
\subsection{Computational details for Stages 1 and 3}
\label{subsec:compute_stage1_3}
Inferential statistical analysis is conducted on the model parameters mentioned in Section \ref{subsec:Stage1} and \ref{subsec:Stage3} based on MCMC sampling. Conjugate priors are selected whenever possible. The full posterior distributions of the model parameters and hyperparameters are provided in the Supplementary Materials. We here briefly outline the MCMC steps. In Stage 2, a fixed rank kriging model was directly fitted using the \texttt{R} package \texttt{FRK}; hence, this was skipped here.


In Stage 1, the parameters and hyperparameters are $\bm{\Theta}_1 = \left\lbrace \lbrace \bm{X}_t \rbrace_{t = 1}^{T}, \bm{\mu}_Z, \bm{\theta}_{\mu}, \tau_{\mu}, \lbrace \bm{\varepsilon}^*_t \rbrace_{t=1}^{T}, \phi_{\varepsilon}, r_{\varepsilon} \right\rbrace$. By an abuse of notation, we reparametrize $\sqrt{r_{\varepsilon}}\bm{\varepsilon}^*_t$ by $\bm{\varepsilon}^*_t$. The full posterior distribution of the latent variables $X_t(\bm{s}_i)$ depends on $Z_t(\bm{s}_i)$. If $Z_t(\bm{s}_i)$ is missing, the posterior distribution of $X_t(\bm{s}_i)$ is normal; otherwise, if $Z_t(\bm{s}_i)$ is zero or one, the posterior of $X_t(\bm{s}_i)$ is a truncated normal distribution, supported on the negative or positive side of the real line, respectively. The prior distribution we choose for $\bm{\mu}_Z$ is $\bm{\mu}_Z \sim \textrm{Normal}_N(\bm{D} \bm{\theta}_{\mu}, \tau^{-1}_{\mu} \bm{I}_N)$. Here, $\bm{D}$ is a $(N\times 6)$-dimensional design matrix with its columns representing an intercept term, longitude, latitude, mean altitude, standard deviation of altitude, and the proportion of a pixel that is within the US Mainland. Because the aim of this study was to predict the underlying spatial process at a new set of locations, it is reasonable to assume an unstructured covariance for the prior of $\bm{\mu}_Z$. For the hyperparameters $\bm{\theta}_{\mu}$ and $\tau_{\mu}$, we choose weakly-informative conjugate priors $\bm{\theta}_{\mu} \sim \textrm{Normal}_6(\bm{0}, 10^2 \bm{I}_6)$ and $\tau_{\mu} \sim \textrm{Gamma}(0.1, 0.1)$. The unconditional distribution of $\bm{\varepsilon}^*_t$ is $\bm{\varepsilon}^*_t \sim \textrm{Normal}_{N^*}(\bm{0},  r_{\varepsilon} \bm{Q}_{\phi_{\varepsilon}}^{-1})$ and the conditional distribution of $\bm{X}_t$ given $\bm{\varepsilon}^*_t$ is $\bm{X}_t \vert \bm{\varepsilon}^*_t \sim \textrm{Normal}_{N}(\bm{\mu}_Z + \bm{A} \bm{\varepsilon}^*_t, (1 - r_{\varepsilon}) \bm{I}_N)$. Thus, the full conditional posterior distribution of $\bm{\varepsilon}^*_t$ is an $N^*$-variate normal distribution; furthermore, the calculation of the mean vector and covariance matrix is straightforward. Based on the remaining parameters and hyperparameters, $\bm{\varepsilon}^*_t; t=1,\ldots, T$, are conditionally independent and thus are updated in parallel. In case of the parameters $\phi_{\varepsilon}$ and $r_{\varepsilon}$, any existence of conjugate priors is not known and hence independent priors are selected as $\phi_{\varepsilon} \sim \textrm{Uniform}(0, 2\Delta_{\mathcal{S}})$ and $r_{\varepsilon} \sim \textrm{Uniform}(0, 1)$, where $\Delta_{\mathcal{S}}$ is the largest Euclidean distance between two data locations. The posterior samples from $\phi_{\varepsilon}$ and $r_{\varepsilon}$ are drawn using the well-established Metropolis-Hastings (M-H) algorithm.

In Stage 3, the parameters and latent variables are $\bm{\Theta}_2 = \left\lbrace \lbrace \bm{\eta}^*_t \rbrace_{t=1}^{T}, \phi_{\eta}, r_{\eta}, \rho_{\eta} \right\rbrace$. The unconditional distribution of $\bm{\eta}^*_t$ is a $2N^*$-variate normal distribution; furthermore, the conditional distribution of $\widehat{\bm{W}}_t$ given $\bm{\eta}^*_t$ is a $2N$-variate normal distribution, (the detailed expressions are presented in Section \ref{subsec:Stage3}). The calculation of the full conditional posterior distribution of $\bm{\eta}^*_t$ is straightforward and it is again a $2N^*$-variate normal distribution. When the remaining parameters and hyperparameters, $\bm{\eta}^*_t; t=1,\ldots, T$, are conditionally independent and thus are updated in parallel. For the parameters $\phi_{\eta}$, $r_{\eta}$, and $\rho_{\eta}$, any existence of conjugate priors is not known and we choose independent priors $\phi_{\eta} \sim \textrm{Uniform}(0, 2\Delta_{\mathcal{S}})$, $r_{\eta} \sim \textrm{Uniform}(0, 1)$, and $\rho_{\eta} \sim \textrm{Uniform}(0, 1)$. Posterior samples from $\phi_{\eta}$, $r_{\eta}$, and $\rho_{\eta}$ are drawn using an M-H algorithm.

Each MCMC chain was run for 60,000  iterations with the first 10,000 iterations discarded as burn-in. The post-burn-in samples were then thinned maintining one in each five samples. Convergence of the chains was monitored via trace plots. The computation of Stage 1 and Stage 3 were undertaken on a desktop with an Intel Xeon CPU E5-2680, a 2.40GHz processor and 128GB RAM , and the corresponding computational times for Stage 1 and Stage 3 were 447 minutes and 744 minutes respectively. Note that these two stages can be run in parallel.

\subsection{Computational details for Stage 4}
\label{subsec:compute_stage4}

We used the function \texttt{randomForest} from the \texttt{R} package \texttt{randomForest}, which implements Breiman's random forest algorithm. The constructed RF models  are trained and their performances tested using a cross-validation study detailed in Section \ref{subsec:CV}. For classification problems, the optimal number of covariates used at each splitting node is $\sqrt{P}$, where $P$ is the total number of covariates. However, multiple models were fitted using different number of covariates, and the model with the best prediction performance was selected based on a cross-validation scheme. The running time on a computer with the same configuration described in Section \ref{subsec:compute_stage1_3}, was approximately 20 minutes.  After some trial-and-error, the tuning parameter configurations were finally chosen as follows: $mtry=36$ (all available covariates), where $mtry$ is the number of covariates randomly sampled as candidates for each split; and $ntree= 200$, where $ntree$ denotes the number of trees to grow. Note that $ntree$ should not be extremely small to ensure that every input row is predicted a few times. These two tuning-parameters are selected based on some exploratory experiments and a cross-validation study.




\subsection{Computational details for fitting the LGCP model}
Posterior inference is obtained from the LGCP model \eqref{eq:LGCP} based on a stochastic gradient-based MCMC method (For more details refer to \cite{welling2011bayesian} and Algorithm 1 in \cite{yadavetal2021}). In LGCP models of the form \eqref{eq:LGCP}, the set of parameters, hyperparameters, and latent variables is given by $\bm{\Theta}_{\textrm{LGCP}}= \left\lbrace \lbrace \bm{\Lambda}_t \rbrace_{t = 1}^{T}, \bm \beta, \lbrace \bm{\zeta}^*_t \rbrace_{t=1}^{T}, \phi_{\zeta}, r_{\zeta} \right\rbrace$, where $\bm \beta$ is the vector of regression coefficients associated with the known covariates, and $\left\lbrace \lbrace \bm{\zeta}^*_t \rbrace_{t=1}^{T}, \phi_{\zeta}, r_{\zeta} \right\rbrace$ has a similar interpretation as $\left\lbrace\lbrace \bm{\varepsilon}^*_t \rbrace_{t=1}^{T}, \phi_{\varepsilon}, r_{\varepsilon} \right\rbrace$ in Section \ref{subsec:Stage1}.  We use conjugate priors whenever possible, and these parameters are updated using Gibbs sampling. For $\bm \beta$, we choose a weakly-informative conjugate Gaussian prior with mean zero and variance 100 when a small number of covariates are in the model, and for the case of a large number of covariates, we use an informative Gaussian prior with mean 0 and variance equal to 0.1 that allows appropriate penalization. For the Mat\'ern correlation parameters $\phi_{\zeta}$ and $r_{\zeta}$, we use the same priors as for $\phi_{\varepsilon}$ and $r_{\varepsilon}$, respectively. We update them within MCMC using a M-H algorithm, similar to updating $\phi_{\varepsilon}$ and $r_{\varepsilon}$. For the latent vectors $\lbrace \bm{\Lambda}_t \rbrace_{t = 1}^{T}$, we do not have closed-from posteriors, and thus we update them jointly using the stochastic gradient Langevin dynamics, which is similar to Algorithm 1 in \cite{yadavetal2021}. For the latent vectors $\lbrace \bm{\zeta}^*_t \rbrace_{t=1}^{T}$, we have closed-form full posteriors, and thus they are updated using Gibbs sampling. For the stochastic gradient MCMC algorithm, we set the batch size to be $b=10$ (i.e., we update 10 out of the $T$ vectors $\bm{\Lambda}_t$ at a time). Because the remaining parameters and hyperparameters, $\bm{\zeta}^*_t; t=1,\ldots, T$, are conditionally independent, we update them in parallel. 

The MCMC chains were run for a total of 250,000 iterations, and the first 200,000 samples were discarded as burn-in samples.  The chains were thinned by keeping one for each 25 samples. Thus, all the summary statistics were calculated based on the final 2000 samples. The computation is approximately 37 hours, when only spatial covariates are included in the models, and approximately 110 hours when we use all the covariates (both spatial and spatio-temporal). 


\section{Data Application}
\label{sec:DataAppl}
\subsection{Cross-validation schemes and model comparison} 
\label{subsec:CV}

The prediction performance of the four-stage model proposed in Sections \ref{subsec:Stage1} to \ref{subsec:Stage4} is compared to a few sub-models, in addition to the alternative LGCP model described in Section \ref{subsec:lgcp}. For comparison, a cross-validation study was performed by dividing the available data after masking into training and test sets. To ensure that the used cross-validation scheme is compatible to that used by the EVA 2021 data challenge organizers, we attempt to replicate the original missingness pattern in the test set, and we choose the same metric of model comparison as that used for the data challenge described in Section \ref{subsec:data_description}.

\begin{figure}[ht]
\centering
\includegraphics[width=1\linewidth]{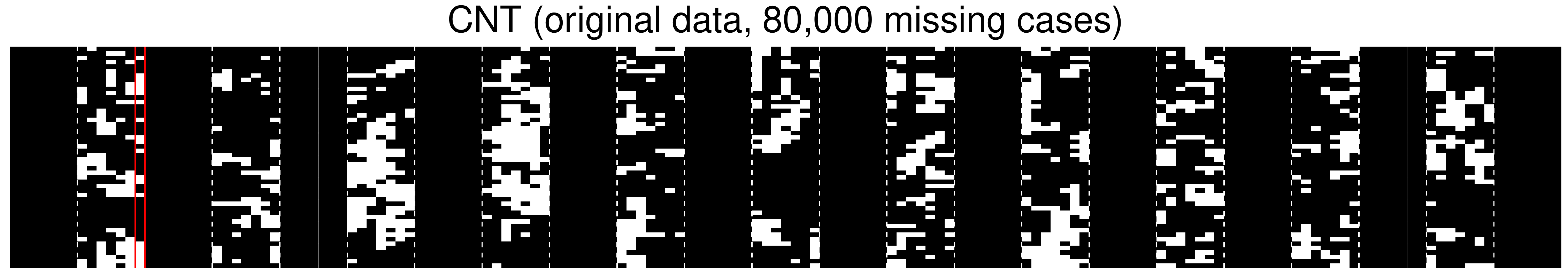}
\includegraphics[width=1\linewidth]{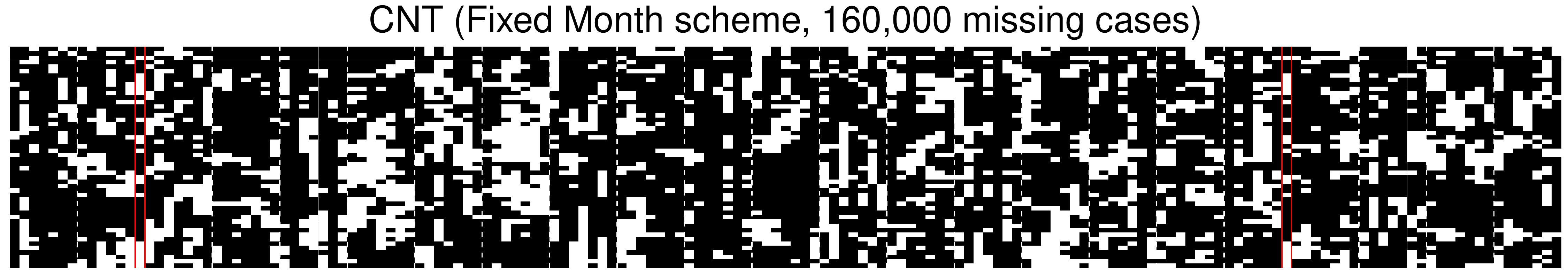}
\caption{A graphical schematic of the cross-validation schemes. The top panel, we illustrates binary time series (across the X-axis) for the first 50 pixels (across the Y-axis), where data availability and missingness are presented in black and white, respectively. The pixels within two consecutive white dashed lines correspond to seven months within a year. The 14th column (September 1994), which highlighted within the red vertical lines, is used in the bottom panel for illustration. The bottom panel, shows the binary time series for data availability and missingness, after masking additional 80,000 spatiotemporal cases under the `Fixed month' scheme. Here, column 133 (September 2011) has the same missingness pattern as column 14.}
\label{fig:cv}
\end{figure}

As reported in Section \ref{subsec:data_description}, the observations are available for 23 years (1993--2015), from March to September. Thus, data for a total of 161 months were available. Of the 23 years, complete observations for 12 years (84 months) are available, and a significant proportion of the data are missing (\texttt{NA}) for the remaining of 11 years (77 months). Here, two types of test sets were developed by replicating the spatial patterns of the missing observations for 77 months out of 84 months (chosen randomly) with complete data. The first type is called `Fixed Month' scheme, where missingness is created for a month using the pattern from the same month index; e.g., for March 1993 (has no missing data), the month March is randomly selected for one out of 11 years where certain data are missing, e.g. 1994, and then the missingness pattern for March 1993 is set to be the same as that for March 1994. The second type is called `Random Month' scheme where missingness for a month is created using the pattern from any randomly selected month with certain observations missing. Figure \ref{fig:cv} shows the process of generating the new test set for CNT using the principle of the `Fixed Month' scheme. The original test set created by the organizers contains 80,000 spatiotemporal locations, and we masked data at additional 80,000 observations. These two types of cross-validation schemes are primarily selected to confirm the validity of the prediction performance of the proposed models while accounting for seasonality and clustering of masked data in space.


For BA prediction, five competing models, closely linked to the final proposed model, are compared. These include the benchmark, which is a linear regression model for BA on a logarithmic scale, with all covariates including all filled CNT (achieved thanks to a Poisson regression). The spatial prediction performances, of the five models are compared using the same metric employed by the data challenge organizers. For the first competitor, a scenario where there is no data available for CNT,  is assumed; thus, the zero/nonzero indicator created in Stage 1 is only based on the BA information. Consequently, the original 80,000 missing observations (rather than 48,947 cases, as mentioned in Section \ref{subsec:data_description}) in Stage 1 are obtained. Moreover, rather than a bivariate  modeling of BA and CNT, a similar univariate SPDE-based spatial model is used for BA only. For other competing models, the CNT data is assumed to be available. For the second competitor, CNT data are used only for the zero/nonzero indicator part in Stage 1, for which data of only 48,947 spatiotemporal locations are missing. However, for the positive part of BA, the same univariate SPDE-based spatial model is fitted as in the case of the first competitor. 

For the third competitor, all the 35 spatiotemporal, purely spatial, or purely temporal covariates mentioned in Section \ref{subsec:data_description} are used in Stage 1. However, the same univariate SPDE-based model in Stage 3 as for the first two competitors, is fitted. In Stage 1, we replace (\ref{z_def}) as follows
\begin{eqnarray}
\nonumber  Z_t(\bm{s}_i) = \begin{cases}
        1, & \parbox[t]{.2\textwidth}{if~~$X_t(\bm{s}_i) > 0$}\\
        0, & \text{if~~$X_t(\bm{s}_i) < 0$},
    \end{cases}
    \textrm{where}~ X_t(\bm{s}_i) = \alpha(\bm{s}_i) + \sum_{l=1}^{35} \gamma_{l} D_{l,t}(\bm{s}_i)+ \varepsilon_t(\bm{s}_i),
\end{eqnarray}
where $\alpha(\cdot)$ is a spatially-varying intercept term, $D_{l,t}(\bm{s}_i), l=1, \ldots, 35$, are the 35 covariates, $\gamma_l, l=1, \ldots, 35$, are the corresponding spatially/temporally invariant regression coefficients, and $\varepsilon_t(\bm{s}_i)$ are the same as in (\ref{z_def}). Here, $\textrm{Pr}(Z_t(\bm{s}_i) = 1) \approx \Phi(\alpha(\bm{s}_i) + \sum_{l=1}^{35} \gamma_{l} D_{l,t}(\bm{s}_i))$. Table \ref{table:cv} lists the evaluation scores for both cross-validation schemes. Under both schemes, the final model described in Stages 1 to 3 exhibited better performance than the alternative univariate spatial models and the benchmark.

\begin{table}[]
\centering
\caption{EVA 2021 data challenge evaluation scores for different models and settings under the two proposed cross-validation schemes, i.e., Fixed Month and Random Month. (A smaller value indicates better performance).}
\label{table:cv}
\begin{tabular}{l | c | c}
\hline
\multicolumn{3}{l}{\textbf{Cross-Validation for BA}}  \\
\hline
Model  & Fixed Month & Random Month \\
\hline
\rowcolor{gray!40}
\begin{tabular}[c]{@{}l@{}}Benchmark model (Log-Gaussian \\ regression, with filled CNT)\end{tabular}                           &       3468.85      &   3574.88           \\
\begin{tabular}[c]{@{}l@{}}Univariate spatial modeling, \\ with BA zero/nonzero indicator\end{tabular}                           &       2842.88      &    2972.47          \\
\rowcolor{gray!40}
\begin{tabular}[c]{@{}l@{}}Univariate spatial modeling, \\ with BA/CNT zero/nonzero indicator\end{tabular}                       &     2842.64        &      2972.25        \\
\begin{tabular}[c]{@{}l@{}}Univariate spatial modeling, \\ with BA/CNT zero/nonzero \\ indicator and all covariates\end{tabular} &       3841.73        &    3971.32       \\
\rowcolor{gray!40}
\begin{tabular}[c]{@{}l@{}}Bivariate spatial modeling, \\ with BA/CNT zero/nonzero indicator\end{tabular}                        &    \textbf{2796.93}         &     \textbf{2923.08}   \\
\hline
\multicolumn{3}{l}{\textcolor{white}{Extra Space}}\\
\hline
\multicolumn{3}{l}{\textbf{Cross-Validation for CNT}} \\
\hline
Model  & Fixed Month & Random Month \\
\hline
\rowcolor{gray!40}
\begin{tabular}[c]{@{}l@{}}Benchmark model (Poisson regression,\\ with log-link, excluding BA)\end{tabular}                           &    5663.02        &     5218.30         \\
\begin{tabular}[c]{@{}l@{}}LGCP with SPDE basis with \\ no covariates (only intercept)\end{tabular}                                               &     4369.05    &      4822.24          \\
\rowcolor{gray!40}
\begin{tabular}[c]{@{}l@{}}LGCP with SPDE basis and \\ purely spatial covariates\end{tabular}                                    &  4358.78           &  4798.53            \\
\begin{tabular}[c]{@{}l@{}}LGCP with SPDE basis and \\ purely spatial covariates and BA\end{tabular}                             &        4469.38     &        4835.74      \\
\rowcolor{gray!40}
\begin{tabular}[c]{@{}l@{}}LGCP with SPDE basis and \\ all covariates and BA\end{tabular}                                        &       4423.26      &         4795.33     \\
\begin{tabular}[c]{@{}l@{}} Bivariate spatial modeling, with BA/CNT \\ zero/nonzero indicator, without Stage 4\end{tabular}  &           3821.24  &  3570.91     \\
\rowcolor{gray!40}
\begin{tabular}[c]{@{}l@{}}RF with \textit{mtry} = 3 and \textit{ntree} = 200\end{tabular}  &    2784.01 & 2990.90        \\
\begin{tabular}[c]{@{}l@{}}RF with \textit{mtry} = 6 and \textit{ntree} = 200\end{tabular}  &   2638.03 & 2827.02      \\
\rowcolor{gray!40}
\begin{tabular}[c]{@{}l@{}}RF with \textit{mtry} = 36 and \textit{ntree} = 200\end{tabular} &    \textbf{2612.01} & \textbf{2785.05}            \\
\hline
\end{tabular}
\end{table}

In CNT prediction, three types of competing models were selected with different settings for each type, including the benchmark, which is a Poisson regression model with all covariates except BA. The first type is a univariate LGCP model constructed using a latent SPDE, as in (\ref{eq:LGCP}). In the first setting, it is assumed that the term $\mu_t(\bm{s}_i)$ in (\ref{eq:LGCP}) is constant across space and time. In the second setting, $\mu_t(\bm{s}_i)$ is assumed to be only spatially-varying and it is written as a linear combination of purely spatial covariates, including an intercept term (the columns of the design matrix $\bm{D}$ in Section \ref{subsec:compute_stage1_3}). In the third setting, $\mu_t(\bm{s}_i)$ is assumed to be both spatially and temporally varying; it is written as a linear combination of the purely spatial covariates, as well as BA, where the missing values are filled using Stages 1 to 3. Finally, in the fourth setting, $\mu_t(\bm{s}_i)$ is written as a linear combination of all covariates and filled BA data. The second model type is the purely statistical model described in Stages 1 to Stage 3, without incorporating any rectification of the erroneous Gaussian assumption for CNT. The third model type is based on RFs, in which all the spatiotemporal covarites, as well as filled BA, are used. Under this setting, we fix $ntree = 200$ as mentioned in Section \ref{subsec:compute_stage4}, and set the number of covariates randomly sampled as candidates for each split (\textit{mtry}) at 3, 6, and 36, under three different settings. The final evaluation scores for the two cross-validation schemes are presented in Table \ref{table:cv}. Under both cross-validation schemes, the final model described in Stages 1 to 4 performs better than the alternative univariate LGCP and the bivariate model without the rectification using RF, as well as the benchmark.


\subsection{Results}

In this section, the posterior means and posterior standard deviations of the model parameters and hyperparameters in Stages 1 and 3 are discussed. The fixed rank kriging estimates of parameter surfaces $\mu_1(\cdot)$, $\mu_2(\cdot)$, $\sigma_1(\cdot)$, and $\sigma_2(\cdot)$ in Stage 2, and the prediction performance of the final model described in Stage 1 through Stage 4 are also reported and discussed.

\begin{figure}[h]
\centering
\includegraphics[width=1\linewidth]{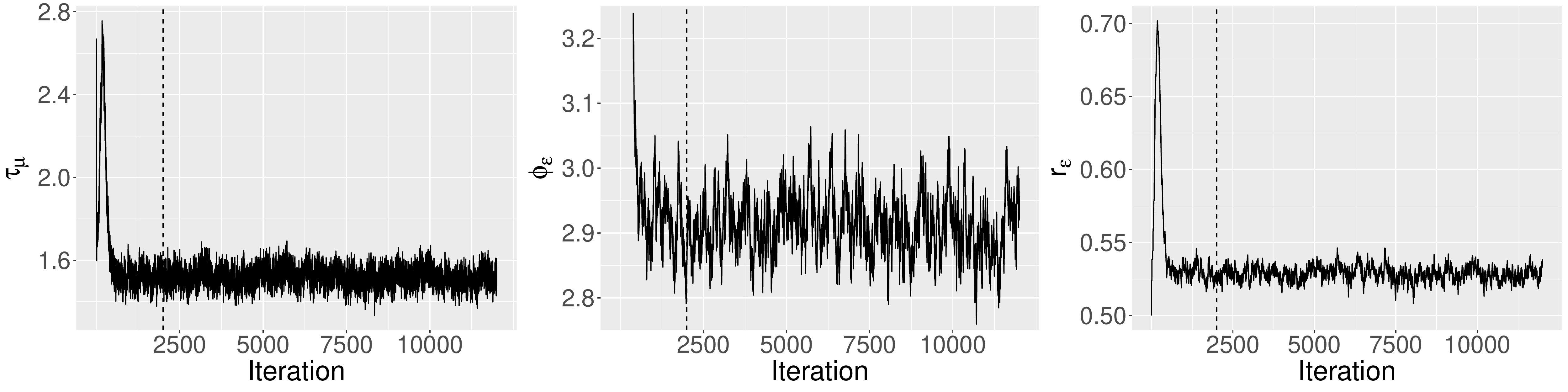}
\includegraphics[width=1\linewidth]{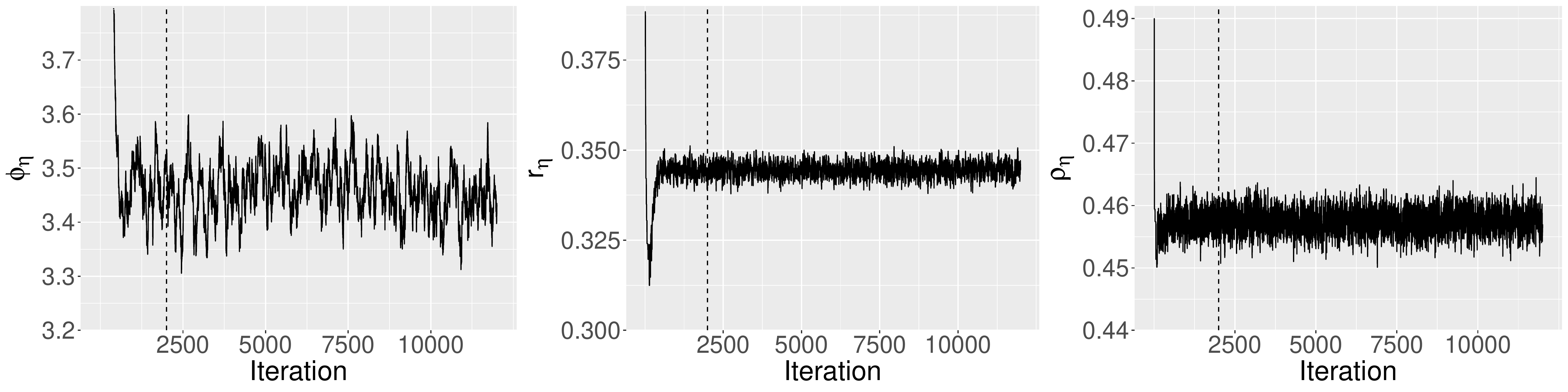}
\caption{Trace plots for some parameters and hyperparameters in Stages 1 (top) and 3 (bottom).}
\label{fig:trace_plots}
\end{figure}

The mixing and convergence of MCMC chains in Stages 1 and 3 were evaluated using trace plots. For scalar parameters, Figure \ref{fig:trace_plots} shows the trace plots of the thinned MCMC chains. All plots exhibit good mixing and convergence diagnostics. For other parameter vectors and latent variables, the convergence and mixing are also confirmed though, not shown here. The computations in Stages 2 and 4 were conducted using the \texttt{R} packages \texttt{FRK} and \texttt{randomForest}, respectively.

\begin{figure}[h]
\centering
\includegraphics[width=1\linewidth]{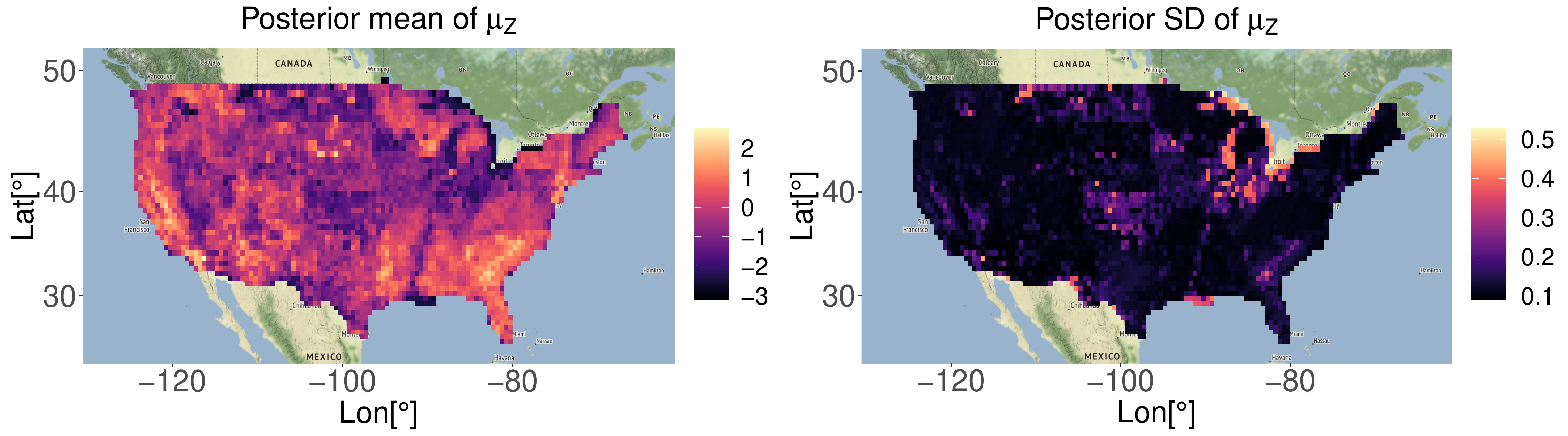}
\caption{Posterior mean and standard deviation profiles of $\mu_Z(\cdot)$.}
\label{fig:posterior_mu}
\end{figure}

Figure \ref{fig:posterior_mu} shows the spatial maps of the posterior mean and posterior standard deviation of $\mu_Z(\cdot)$ in Stage 1. The values of the posterior mean of $\mu_Z(\cdot)$ are higher near the Southeastern (the state of Georgia) and Southwestern (the state of California) parts of the US and  are generally lower in the middle parts of the US. Among available temporal replications, the posterior standard deviation is extremely high (more than 0.4) for 54 spatial locations, mostly in the northeastern parts of the US, where the zero/nonzero indicator is either always zero (for 53 locations) or always one (for 1 location).

Table \ref{table:posterior_mean_sd} lists the posterior mean and standard deviation of non-spatial parameters and hyperparameters in Stages 1 and 3. In a frequentist sense, all parameters $\theta_{\mu, 1}$ to $\theta_{\mu, 6}$ are significant (the absolute value of the ratio of posterior mean and posterior standard deviation is larger than 2 for each of the six cases). The posterior means of $\phi_{\varepsilon}$ and $r_{\varepsilon}$ are 3.0491 and 0.5319, respectively, for the latent Gaussian process, which indicates a correlation of 0.050 at a spatial distance of 10 degrees. In Stage 3, the posterior means of $\phi_{\eta}$ and $r_{\eta}$ are 3.6408 and 0.3442, respectively, for each component of the bivariate spatial Gaussian process, which indicates a correlation of 0.052 at a spatial distance of 10 degrees. These results show the requirement for modeling the spatial dependence in fire occurrences and sizes. The posterior mean of $\rho_{\eta}$ is 0.4575, which shows a strong positive correlation between BA and CNT.


\begin{table}[ht]
\centering
\caption{Posterior mean and standard deviation (SD) of the non-spatial parameters and hyperparameters in Stage 1 and Stage 3. Here, $\theta_{\mu, 1},\ldots,\theta_{\mu, 6}$ denote the six components of $\bm{\theta}_{\mu}$ that represent an intercept term and the regression coefficients of longitude, latitude, mean altitude, standard deviation of altitude, and the proportion of a pixel that is within the mainland US, respectively.}
\label{table:posterior_mean_sd}
\begin{tabular}{ccc}
  \hline
 & Stage 1 & \\
 \hline
Parameter & Posterior mean & Posterior SD \\ 
\hline
$\theta_{\mu,1}$ & -0.3805 & 0.0262 \\ 
$\theta_{\mu,2}$ & 0.0894 & 0.0239 \\ 
$\theta_{\mu,3}$ & -0.2596 & 0.0192 \\ 
$\theta_{\mu,4}$ & -0.2583 & 0.0280 \\ 
$\theta_{\mu,5}$ & 0.4540 & 0.0289 \\ 
$\theta_{\mu,6}$ & 0.1826 & 0.0154 \\ 
$\tau_\mu$ & 1.5398 & 0.1276 \\ 
$\phi_{\varepsilon}$ & 3.0491 & 0.9754 \\ 
$r_{\varepsilon}$ & 0.5319 & 0.0194 \\ 
\hline
 & Stage 3 & \\
 \hline
Parameter & Posterior mean & Posterior SD \\ 
\hline
$\phi_{\eta}$ & 3.6408 & 1.2179 \\ 
$r_{\eta}$ & 0.3442 & 0.0042 \\ 
$\rho_{\eta}$ & 0.4575 & 0.0022 \\ 
   \hline
\end{tabular}
\end{table}

\begin{figure}[h]
\centering
\includegraphics[width=1\linewidth]{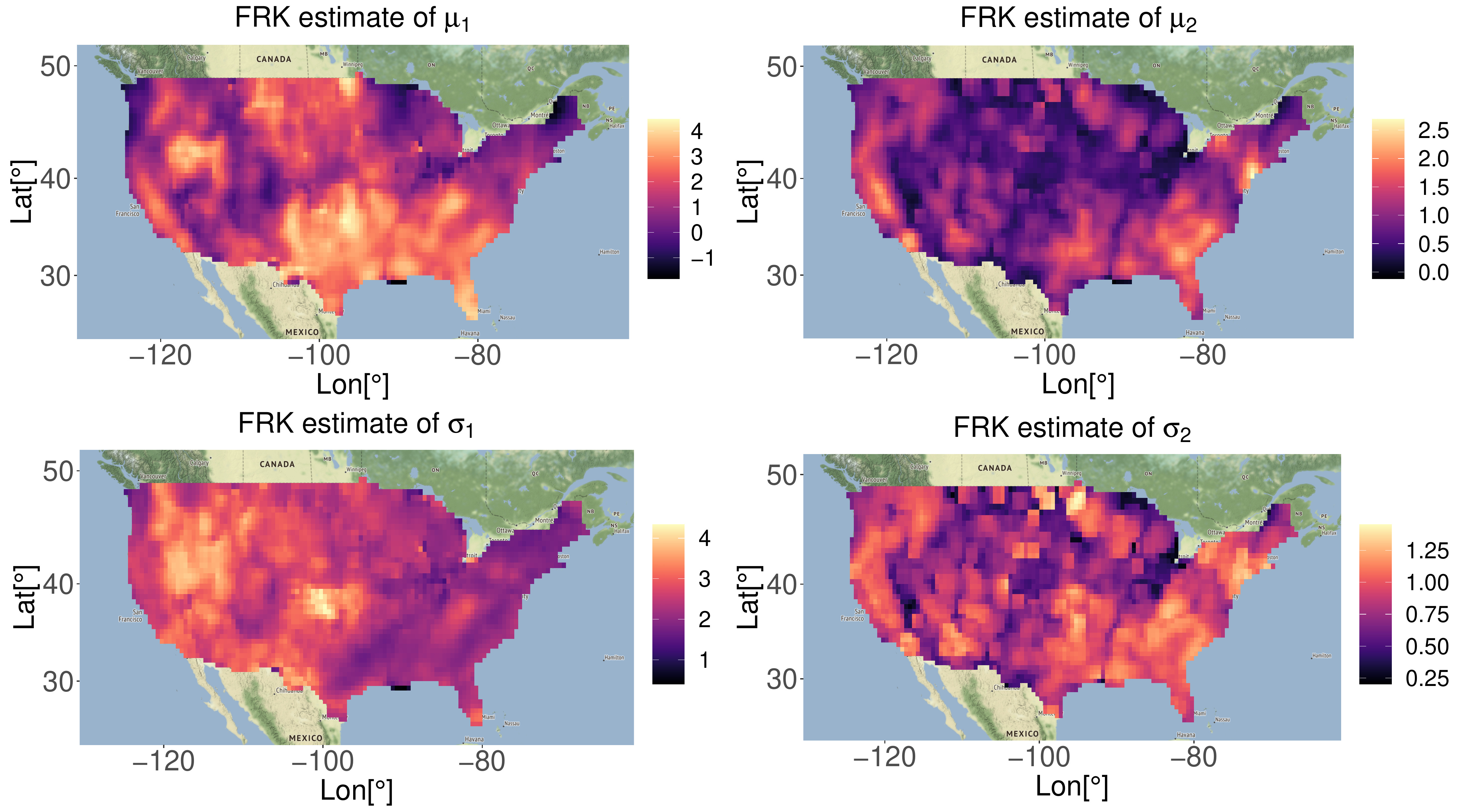}
\caption{Approximate Bayesian estimates of $\mu_1(\cdot)$, $\mu_2(\cdot)$, $\sigma_1(\cdot)$, and $\sigma_2(\cdot)$ using the fixed rank kriging procedure described in Section \ref{subsec:Stage2}.}
\label{fig:posterior_mu_sigma}
\end{figure}

Figure \ref{fig:posterior_mu_sigma} shows the fixed rank kriging estimates of $\mu_1(\cdot)$, $\mu_2(\cdot)$, $\sigma_1(\cdot)$, and $\sigma_2(\cdot)$, in Stage 2. For $\mu_1(\cdot)$, the FRK estimates are generally higher in a large portion of the southeastern US, while the values are lower in the mid-west, north-west and the north-east. For $\mu_2(\cdot)$, the values are higher in two small regions in the southeast and southwest of the US. In some regions (mid-north), the estimated profile is not highly smooth indicating high local nonstationarity. For $\sigma_1(\cdot)$, large estimates are visible in a small portion of the mid-US, while moderately large values are observed in a large region in the western US. For $\sigma_2(\cdot)$, large values are observed in two large regions of the eastern and western US. In the mid-US, the estimated profile is not smooth in some regions, particularly, near the mid-north of the US, similar to the $\mu_2(\cdot)$ profile. The spatial maps of $\sigma_1(\cdot)$ and $\sigma_2(\cdot)$ illustrate the underlying spatial heteroscedasticity and sharp local variability. 

\begin{figure}[ht]
\centering
\adjincludegraphics[width = \linewidth, trim = {{.0\width} {.0\width} {.0\width} {.0\width}}, clip]{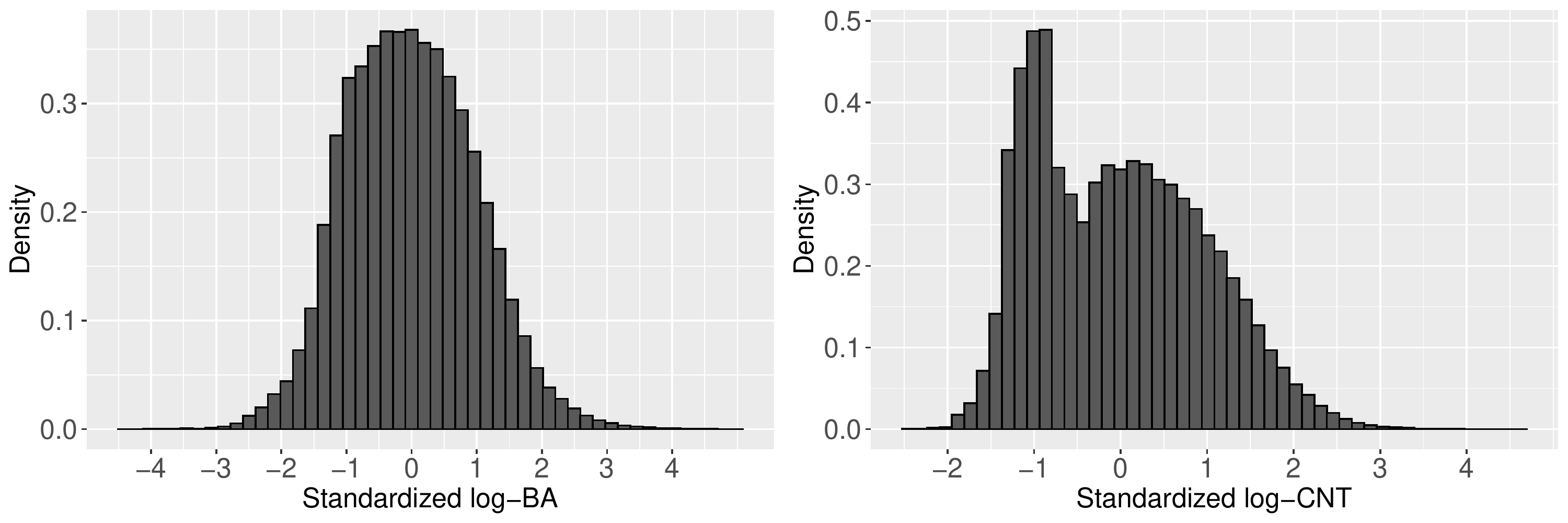}
\caption{Histograms of the standardizing $\log$-BA and $\log$-CNT based on the Stage 2 estimates.}
\label{fig:lack_of_fit}
\end{figure}

After incorporating the Stage 2 estimates, we obtain the residuals $\widehat{W}_{tp}(\bm{s}_i)$ (Section \ref{subsec:Stage3}). Despite fitting a Gaussian model to log-BA and log-CNT in Stage 3 for computational suitability, we study the histograms of the standardized log-BA and log-CNT values to check the validity of our Gaussian assumption, and we present them in Figure \ref{fig:lack_of_fit}. For the standardized log-BA, the histogram is bell-shaped and symmetric around zero and the Gaussian assumption appears to be reasonable. However, for the standardized log-CNT, the histogram is bimodal and the shapes of the two modes are considerably different. Furthermore, the histogram appears to be right-skewed; thus, a Gaussian assumption for $\widehat{W}_{tp}(\bm{s}_i)$ in Stage 3 for CNT is questionable and this justifies the rectification using RFs in Stage 4.

In Stage 4, the RF algorithm, is used. In this algorithm, the final model based on cross-validation includes all the covariates and BA (missing values are imputed using Stages 1--3), as described in Section \ref{subsec:CV}. 

\begin{figure}
   \includegraphics[width=1\linewidth]{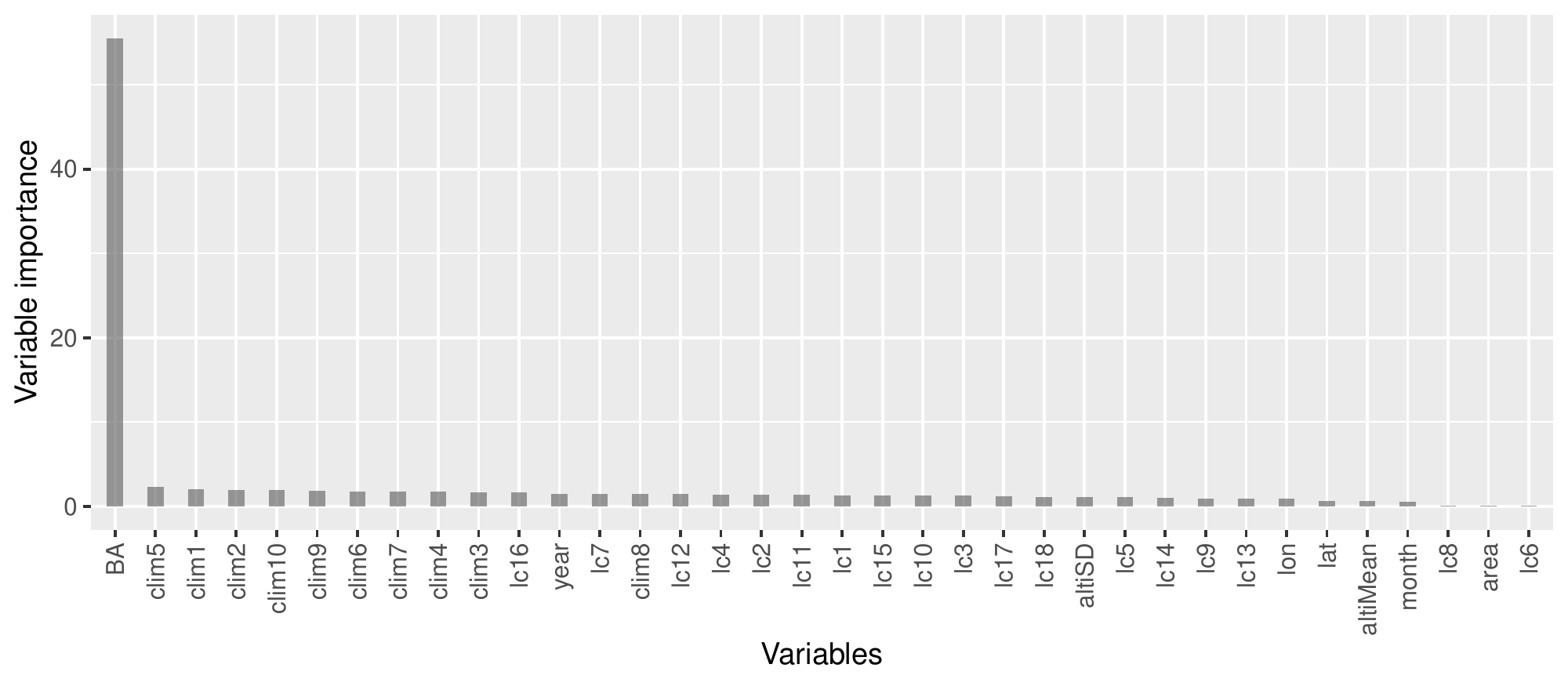}
   \caption{Variable importance (or mean decrease accuracy plot) for the RF algorithm in Stage 4, in which all the available covariates and filled BA were used for classification.}
    \label{fig:var_importance}
\end{figure}

Figure \ref{fig:var_importance} shows the variable importance plot (or mean decrease accuracy plot). The plot expresses how much accuracy the model looses by excluding each variable at a time during classification. The mean decrease in Gini coefficient is a measure of how each variable contributes to the homogeneity of nodes and leaves in the resulting RF. The increase in the value mean reduction in accuracy or mean reduction Gini score, increases the importance of the variable in the model. Here, BA is reported to be the most important covariate, followed by clim5 (Evaporation) and clim1 (wind speed in Eastern direction), whereas lc6 (tree broadleaved deciduous closed) is the least important covariate. The covariates were included into the models in the order of their variable importance.

\begin{figure}[ht]
\centering
\adjincludegraphics[width = \linewidth, trim = {{.0\width} {.0\width} {.0\width} {.0\width}}, clip]{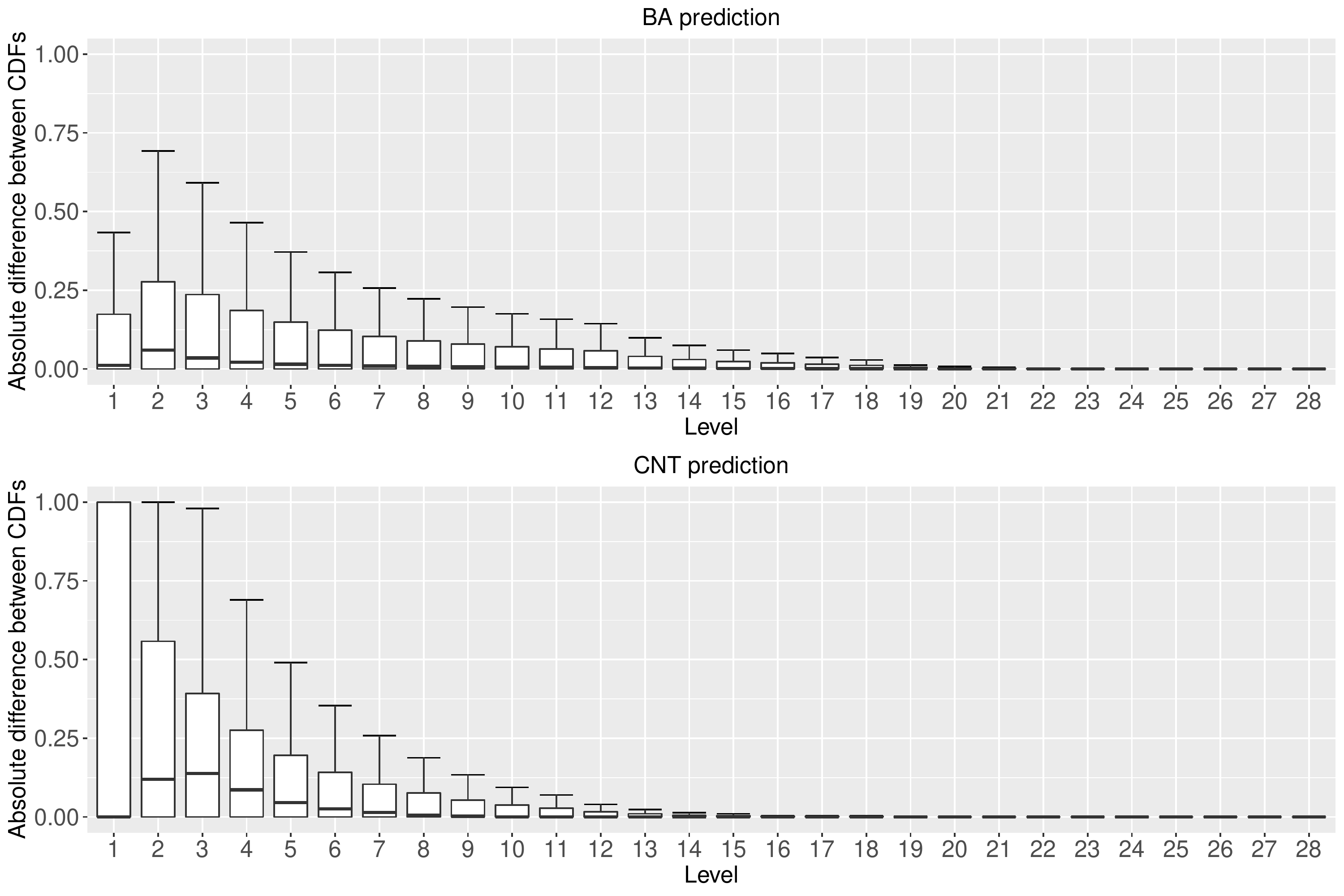}
\caption{Boxplots of absolute differences between the estimated predictive distribution functions and empirical distribution functions evaluated at 28 levels given in (\ref{severity_thresholds}).}
\label{fig:boxplot_cdf_diff}
\end{figure}

Finally, we compare the estimated predictive distribution functions for the 80,000 test observations, for each BA and CNT, evaluated at 28 levels, with the empirical distribution functions calculated based on the test observations. Figure \ref{fig:boxplot_cdf_diff} shows boxplots of the absolute differences between the two cumulative distribution functions (CDFs) at 28 evaluation levels, specified in (\ref{severity_thresholds}). At lower levels certain boxes and upper endpoints of error bars are significantly different from zero; however,  the median of the absolute differences is close to zero in most cases, particularly for BA. For higher levels, which are assigned larger weights for model evaluation, the boxes are close to zero indicating that the proposed model performs well for predicting the distribution of BA and CNT at unobserved spatiotemporal locations.

\section{Discussions and Conclusions}
\label{sec:Conclusions}

\subsection{Summary}

Motivated by the EVA 2021 data challenge, in which the first four authors of this paper participated as the \textit{The Bedouins} team, a four-stage high-dimensional zero-inflated bivariate spatial model based on statistics and machine learning was proposed for the prediction of BA and CNT at masked spatiotemporal locations. Here, a spatial dependence structure was developed using SPDEs, which reduces the computational burden by allowing sparsity in the precision matrices. In Stage 1, the data were categorized into zero/nonzero categories and a two-layered Bayesian hierarchical model was fitted for estimating the probabilities of the two categories at unobserved sites. In Stage 2, the parameter surfaces were estimated using two-step approximate Bayesian inference technique that bypasses high computational burden.

In Stage 3, the standardized log-transformed positive observations from Stage 2 were modeled using a bivariate spatial GMRF. The log-Gaussian assumption for modeling positive wildfire frequencies was computationally helpful; however, it was erroneous because the observations are discrete-valued. Thus, in Stage 4, the predicted values of wildfire frequencies were rectified using RFs, in which BA was treated as a covariate, and the missing BA values were imputed by the predicted BA values in Stage 3.  MCMC sampling was used to draw posterior inference in Stages 1 and 3. The computation in Stages 2 and 4 were done directly using the \texttt{R} packages \texttt{FRK} and \texttt{randomForest}, respectively. Our final model was shown to outperform some alternatives in a well-designed cross-validation study and to effectively predict low to high quantiles of BA and CNT at unobserved sites. Table \ref{table:final_scores} lists the final scores for the three best participating teams. While the teams \textit{BlackBox} and \textit{Kohrrelation} focused on purely ML-based approaches, our method only depends on ML algorithms for rectifying predictions given by statistical models.
 
  A cross-validation scheme was created to effectively compare the models. For modeling discrete spatial data, an alternative approach using LGCPs was discussed. The fitted LGCP was constructed using a latent SPDE spatial effect; however, in case of spatial prediction of the US wildfire data, our proposed model outperforms the LGCP approach. While fully machine learning algorithms do not have any distributional assumption and are thus more robust, our proposed statistical method can better quantify the underlying uncertainty, specifically in Stages 1 and 3.


\begin{table}[]
\centering
\caption{Final evaluation scores for the three best performing teams in the EVA 2021 data challenge and the benchmark. (A smaller score indicates better performance).}
\label{table:final_scores}
\begin{tabular}{c| ccc c}
  & BlackBox & Kohrrelation & The Bedouins & Benchmark \\
  \hline 
BA                           & 3315.65  & 3446.02      & 3408.31 & 4244.36 \\
CNT                          & 2804.95  & 2989.85      & 3145.81 & 5565.15 \\
\hline 
Total                        & 6120.60  & 6435.87      & 6554.12 & 9809.51 \\
\hline
\end{tabular}
\end{table}


\subsection{Drawbacks and possible solutions}

Despite its elegant performance in predicting the distribution functions at masked sites, our approach also has some limitations. First, we selected a multistage approach in which information was not borrowed from one stage to another. For example, in Stage 1, the data are treated as zero/nonzero indicators; however, in other stages, zeros are treated as missing data. Joint modeling of zero and nonzero data is a possible solution. Second, in Stages 1 and 3, it was assumed that the underlying dependence structures are of the isotropic Mat\'ern form given in (\ref{cov_structure}). Because the spatial domain is large, it would be reasonable to assume a nonstationary dependence structure, e.g., using empirical orthogonal functions \citep{wikle2010low, hazra2021estimating}, that are more realistic for modeling the wildfire process over the entire US Mainland. In Stage 2, we obtained smoothed parameter estimates in two steps;  in the first step, we calculated empirical estimates, and in the second step, we treated such estimates as spatial data. Moving from Step 1 to Step 2, we ignored the uncertainty of the estimates in the first step. While the posterior coverage probabilities are only mildly affected by such a choice as shown in \cite{hazra2019spatio}, a superior method for borrowing the uncertainty information from the first step to the second step has been recently discussed in \cite{hrafnkelsson2021max}. In Stage 4, we used the univariate random forest algorithm of \cite{breiman2001random} independently at each site. However, recently, a random forest algorithm for spatially-dependent data has been proposed by \cite{saha2021random}. Because we ignored covariate information in Stage 1 to Stage 3, the proposed model is not suitable to draw inference about the significance of a specific predictor in the context of fire modeling. A possible solution would be to fit a regression model for the mean surfaces. 

\subsection{Other applications}
The model proposed here is a general tool for all bivariate zero-inflated spatiotemporal datasets, although our methodology has been motivated by the joint analysis of fire occurrences and sizes. The model supplements spatial marked point processes; thus, it is suitable for multiple scenarios. For example, \cite{penttinen1992marked} discussed a marked point process approach for forest statistics, where `points' are the tree positions, whereas the `marks' are tree characteristics such as stem diameters and tree species. After summarizing the data over a grid, we can obtain a bivariate spatial dataset with two components representing the number of trees and average stem diameter within each grid cell. In such a scenario, our proposed inference approach would be suitable. Another example is the joint modeling of the number of rainy days within a month and the total monthly precipitation within a pixel/region. In addition to the marked point process scenario, our model can also be used to fit temperature and rainfall data over a spatiotemporal domain \citep{gelfand2005spatial}, for example. Temperature is a real-valued continuous variable for which a Gaussian assumption is common (and thus a log-transformation is not necessary), and rainfall data are non-negative and zeros for the dry periods; here, we can ignore Stage 4 if the log-transformed nonzero precipitation amounts follow a normal distribution.

\subsection{Possible extensions}

The method proposed in Sections \ref{subsec:Stage1} and  \ref{subsec:Stage4} can be extended in multiple directions. The distribution function estimation problem for the EVA 2021 data challenge can be treated as a classification problem. Thus, by categorizing the data based on the 28 levels,  a set of spatial indicators can be obtained, as described in \cite{agarwal2021copula}. Furthermore, a multivariate version of Stage 1 would be a reasonable model for drawing inference. The proposed model assumes temporal independence; thus, a natural extension would be to assume that the latent processes have temporal autocorrelation. By replacing the mean profiles as functions of the covariates, we can extend the proposed model to study the significance of a specific predictor in the context of fire modeling. The underlying dependence structures are assumed to be stationary for Stage 1 and Stage 3; however, some approaches available in the literature allow nonstationary spatial modeling for large datasets \citep{katzfuss2013bayesian, banerjee2020modeling}, and we can extend the proposed model in that direction. Although the main focus in the data challenge is the accurate estimation of the upper tail of CNT and BA observations, we build our model using Gaussian processes that have been criticized for modeling spatial extremes \citep{davison2013geostatistics}. While the extreme-value theory-justified models typically entail a huge computational burden even in low dimensions, a simpler alternative would be to consider scale mixture models \citep{huser2017bridging, huser2020advances, hazra2021spatialscale}. Although we used an adaptive Metropolis-Hastings algorithm for the computations in Stages 1 and 3, we can extend it using some faster and more recently developed algorithms, like the stochastic gradient-based \citep{welling2011bayesian} or deterministic transformation-based \citep{dutta2014markov} MCMC algorithms.


\section*{Acknowledgments}

The first three authors (Cisneros, Gong, Yadav) contributed equally to this work by implementing some of the methods and writing parts of the paper. The last two authors (Hazra, Huser) oversaw the whole project, with Hazra having a leading role throughout all practical aspects of the data competition (supervision of the Bedouins Team, methods' implementation, results' interpretation, writing). 

We would like to thank Thomas Opitz for organizing this very interesting data competition for the EVA 2021 Conference, as well as Thomas Mikosch for welcoming a Special Issue in Extremes about this topic. This publication is based upon work supported by the King Abdullah University of Science and Technology (KAUST) Office of Sponsored Research (OSR) under Award No. OSR-CRG2020-4394.



\bibliographystyle{apalike}
\bibliography{sn-bibliography}


\newpage

\begin{center} {\Large{\bf Supplementary Material}}
\end{center}

\baselineskip=12pt
\vskip 5mm


\baselineskip=17pt
\vskip 4mm
\vskip 6mm

\baselineskip=16pt

\baselineskip=25pt

\section{MCMC details}

\subsection{Stage 1: Spatial indicator kriging for wildfire occurrence data}
We define the indicator variables $Z_t(\bm s_i)$ as
\begin{align*}
  Z_t(\bm{s}_i) = \begin{cases}
  1, \hspace{.4cm}\textrm{if}\begin{cases}
  \parbox[t]{.8\textwidth}{$\textrm{BA}_t(\bm{s}_i)>0, ~\textrm{CNT}_t(\bm{s}_i) > 0$,}\\
    \text{$\textrm{BA}_t(\bm{s}_i)>0, \textrm{CNT}_t(\bm{s}_i)$ is missing, $~$\textrm{or}$~~$}
  \text{$\textrm{CNT}_t(\bm{s}_i) > 0, \textrm{BA}_t(\bm{s}_i)$ is missing,}
  \end{cases}\\ \vspace{-5mm}  \\
  0, \hspace{.4cm}\textrm{if}\begin{cases}
      \parbox[t]{.8\textwidth}{$\textrm{BA}_t(\bm{s}_i)=0, ~\textrm{CNT}_t(\bm{s}_i) = 0$,}\\
      \text{$\textrm{BA}_t(\bm{s}_i)=0, \textrm{CNT}_t(\bm{s}_i)$ is missing, $~$\textrm{or}$~~$}
  \text{$\textrm{CNT}_t(\bm{s}_i)=0, \textrm{BA}_t(\bm{s}_i)$ is missing,}
      \end{cases}\\
    $\texttt{NA},$~~ \text{if ~~$\textrm{BA}_t(\bm{s}_i)$ and $\textrm{CNT}_t(\bm{s}_i)$ both are missing.}   
    \end{cases}
\end{align*}

Further, we assume that $Z_t(\bm{s}_i)$ are independent and identically distributed ($\emph{iid}$) across $t$, and model it using a latent Gaussian process (GP) $X_t(\cdot)$ as
\begin{align*}
    Z_t(\bm{s}_i) = \begin{cases}
        1 & \parbox[t]{.2\textwidth}{if~~$X_t(\bm{s}_i) > 0$}\\
        0 & \text{if~~$X_t(\bm{s}_i) < 0$},
    \end{cases}
    \textrm{where}~ X_t(\bm{s}_i) = \mu_Z(\bm{s}_i) + \varepsilon_t(\bm{s}_i),
\end{align*}
    and $\varepsilon_t(\cdot)$ are spatial GPs that are $\emph{iid}$ across $t$, with $\textrm{E}[\varepsilon_t(\bm{s})] = 0$ and $\textrm{Var}[\varepsilon_t(\bm{s})] = 1$ (approximately). We assume that the process $\varepsilon_t(\cdot)$ follows an isotropic Mat\'ern spatial correlation with range parameter $\phi_{\varepsilon}$ and smoothness parameter fixed at one. Further, we model $\bm{\varepsilon}_t = [\varepsilon_t(\bm{s}_1), \ldots, \varepsilon_t(\bm{s}_N)]'$ as $ \bm{\varepsilon}_t \sim \textrm{Normal}_{N}(\bm{A} \bm{\varepsilon}^*_t, (1-r_{\varepsilon}) \bm{I}_N)$ where $\bm{\varepsilon}^*_t \sim \textrm{Normal}_{N}(\bm{0}, r_{\varepsilon} \bm{Q}_{\phi_{\varepsilon}}^{-1})$. We denote $\bm{\mu}_Z = [\mu_Z(\bm{s}_1), \ldots, \mu_Z(\bm{s}_N)]'$ and choose the prior $\bm{\mu}_Z \sim \textrm{Normal}_{N}(\bm{D} \bm{\theta}_{\mu}, \tau^{-1}_{\mu} \bm{I}_N)$. Also, we denote $\bm{X}_t = [X_t(\bm{s}_1), \ldots, X_t(\bm{s}_N)]'$. The set of latent processes, parameters and hyper-parameters in the model are
\begin{eqnarray}
\nonumber && \bm{\Theta}_1 = \left\lbrace \lbrace \bm{X}_t \rbrace_{t = 1}^{T}, \bm{\mu}_Z, \bm{\theta}_{\mu}, \tau_{\mu}, \lbrace \bm{\varepsilon}^*_t \rbrace_{t=1}^{T}, \phi_{\varepsilon}, r_{\varepsilon} \right\rbrace.
\end{eqnarray}

The MCMC steps for updating the parameters in $\bm{\Theta}_1$ are as follows. For a specific a parameter (or a set of parameters), we call ``$rest$'' the observed data, all the parameters and hyper-parameters in $\bm{\Theta}_1$ except that parameter (or that set of parameters).

Throughout the algorithm, we update the parameters and hyper-parameters consecutively following the full conditional posterior distributions at each iteration as follows:

\noindent \underline{\textbf{$\bm{\mu}_Z | rest$}} \\
The prior distribution of $\bm{\mu}_Z$ is $\bm{\mu}_Z \sim \textrm{Normal}_N(\bm{D} \bm{\theta}_{\mu}, \tau^{-1}_{\mu} \bm{I}_N)$. The full conditional posterior distribution of $\bm{\mu}_Z$ is
\begin{eqnarray}
\nonumber \bm\mu_Z | rest \sim \textrm{Normal}_N\left( \left[ \ffrac{T}{1 - r_{\varepsilon}} + \tau_{\mu} \right]^{-1}  \left[ \ffrac{1}{1 - r_{\varepsilon}} \sum_{t = 1}^{T} (\bm{X}_t - \bm{A} \bm{\varepsilon}^*_t) + \tau_{\mu} \bm{D} \bm{\theta}_{\mu} \right]^{-1}, \left[ \ffrac{T}{1 - r_{\varepsilon}} + \tau_{\mu} \right]^{-1} \bm{I}_N \right).
\end{eqnarray}

\vspace{2mm}

\noindent \underline{\textbf{$\bm{\theta}_{\mu} | rest$}} \\
We consider the conjugate noninformative prior $\bm{\theta}_{\mu} \sim \textrm{Normal}_P(\bm{0}, 10^2 \bm{I}_P)$. The full conditional posterior distribution is
\begin{eqnarray}
\nonumber \bm{\theta}_{\mu} | rest \sim \textrm{Normal}_P\left(\left[\tau_{\mu} \bm{D}'\bm{D} + 10^{-2} \bm{I}_P \right]^{-1} \tau_{\mu} \bm{D}' \bm{\mu}_Z, ~ \left[\tau_{\mu} \bm{D}'\bm{D} + 10^{-2} \bm{I}_P \right]^{-1}\right).
\end{eqnarray}



\noindent \underline{\textbf{$\tau_{\mu} | rest$}} \\
We consider the prior $\tau_{\mu} \sim \textrm{Gamma}(0.1, 0.1)$. The full conditional posterior is
\begin{eqnarray}
\nonumber && \tau_{\mu} | rest \sim \textrm{Gamma}\left(0.1 + N/2, ~ 0.1 + (\bm{\mu}_Z - \bm{D} \bm{\theta}_{\mu})' (\bm{\mu}_Z - \bm{D} \bm{\theta}_{\mu})/2 \right).
\end{eqnarray}

\vspace{2mm}

\noindent \underline{\textbf{$\lbrace\bm{\varepsilon}^*_t\rbrace_{t=1}^{T} | rest$}} \\
The unconditional distribution of $\bm{\varepsilon}^*_t$ is $\bm{\varepsilon}^*_t \sim \textrm{Normal}_{N^*}(\bm{0},  r_{\varepsilon} \bm{Q}_{\phi_{\varepsilon}}^{-1})$. The full conditional posterior distribution of $\bm{\varepsilon}^*_t$ is $\bm{\varepsilon}^*_t | rest \sim \textrm{Normal}_{N^*}(\bm{\mu}^\ast_{\varepsilon^*}, \bm{\Sigma}^\ast_{\varepsilon^*})$, where
\begin{eqnarray}
\nonumber && \bm{\Sigma}^\ast_{\varepsilon^*} = \left[(1 - r_{\varepsilon})^{-1} \bm{A}'\bm{A} + r_{\varepsilon}^{-1} \bm{Q}_{\phi_{\varepsilon}} \right]^{-1}, ~~ \bm{\mu}^\ast_{\varepsilon^*} =  \bm{\Sigma}^\ast_{\varepsilon^*} [(1 - r_{\varepsilon})^{-1} \bm{A}' (\bm{X}_t - \bm{\mu}_Z)].
\end{eqnarray}
The observations are assumed to be independent across $t=1,\ldots, T$, and hence, the full conditional posterior distributions of $\bm{\varepsilon}^*_t; t=1,\ldots, T$ are independent. Thus, $\bm{\varepsilon}^*_t; t=1,\ldots, T$ are updated within MCMC in parallel.

\vspace{2mm}

\noindent \underline{$\phi_{\varepsilon} | rest$} \\
We consider the prior $\phi_{\varepsilon} \sim \textrm{Uniform}(0, 2\Delta_{\mathcal{S}})$, where $\Delta_{\mathcal{S}}$ is the largest Euclidean distance between two data locations. Let the $m$-th MCMC sample from $\phi_{\varepsilon}$ be denoted by $\phi_{\varepsilon}^{(m)}$. Considering a logit transformation, we obtain $\phi_{\varepsilon}^{*(m)} \in \mathbb{R}$ from $\phi_{\varepsilon}^{(m)}$, and simulate $\phi_{\varepsilon}^{*(c)}$ from $\textrm{Normal}(\phi_{\varepsilon}^{*(m)}, s_{\phi_{\varepsilon}}^2)$ distribution, where $s_{\phi_{\varepsilon}}$ is the standard deviation of the candidate normal distribution. Subsequently, using an inverse-logit transformation, we obtain $\phi_{\varepsilon}^{(c)}$ from $\phi_{\varepsilon}^{*(c)}$, and either accept $\phi_{\varepsilon}^{(c)}$ as a candidate from the posterior distribution of $\phi_{\varepsilon}$ or reject it. The acceptance ratio is 
\begin{eqnarray}
\nonumber \mathcal{R} &=& \ffrac{ \prod_{t=1}^{T} f_{\textrm{Normal}_{N^*}}\left(\bm{\varepsilon}^*_t; \bm{0}, r_{\varepsilon} \bm{Q}_{\phi_{\varepsilon}^{(c)}}^{-1} \right) }{ \prod_{t = 1}^{T}  f_{\textrm{Normal}_{N^*}}\left(\bm{\varepsilon}^*_t; \bm{0}, r_{\varepsilon} \bm{Q}_{\phi_{\varepsilon}^{(m)}}^{-1} \right)} \times \ffrac{\phi_{\varepsilon}^{(c)} \left(2\Delta_{\mathcal{S}} - \phi_{\varepsilon}^{(c)}\right)}{\phi_{\varepsilon}^{(m)} \left(2\Delta_{\mathcal{S}} - \phi_{\varepsilon}^{(m)}\right)},
\end{eqnarray}
where $f_{\textrm{Normal}_n}(\cdot; \bm{\mu}, \bm{\Sigma})$ denotes the $n$-variate normal density with mean $\bm{\mu}$ and covariance matrix $\bm{\Sigma}$. The candidate is accepted with probability $min \lbrace \mathcal{R},1 \rbrace$.

\vspace{2mm}


\noindent \underline{$r_{\varepsilon} | rest$} \\
We consider the prior $r_{\varepsilon} \sim \textrm{Uniform}(0, 1)$. Suppose $r_{\varepsilon}^{(m)}$ denotes the $m$-th MCMC sample from $r_{\varepsilon}$. We simulate a candidate sample $r_{\varepsilon}^{(c)}$ from $r_{\varepsilon}^{(m)}$ following a procedure similar to simulating $\phi_{\varepsilon}^{(c)}$ from $\phi_{\varepsilon}^{(m)}$. The Metropolis-Hastings acceptance ratio is 
\begin{eqnarray}
\nonumber \mathcal{R} = \ffrac{ \prod_{t=1}^{T}  f_{\textrm{Normal}_N}\left(\bm{X}_t; \bm{\mu} + \bm{A} \bm{\varepsilon}^*_t, (1 - r^{(c)}) \bm{I}_N \right)}{ \prod_{t = 1}^{T}  f_{\textrm{Normal}_N}\left(\bm{X}_t; \bm{\mu} + \bm{A} \bm{\varepsilon}^*_t, (1 - r^{(m)}) \bm{I}_N \right)}  \times \ffrac{ \prod_{t=1}^{T} f_{\textrm{Normal}_{N^*}}\left(\bm{\varepsilon}^*_t; \bm{0}, r_{\varepsilon}^{(c)} \bm{Q}_{\phi_{\varepsilon}}^{-1} \right) }{ \prod_{t = 1}^{T}  f_{\textrm{Normal}_{N^*}}\left(\bm{\varepsilon}^*_t; \bm{0}, r_{\varepsilon}^{(m)} \bm{Q}_{\phi_{\varepsilon}}^{-1} \right)} \times \ffrac{r_{\varepsilon}^{(c)}  \left(1 - r_{\varepsilon}^{(c)}\right)}{r_{\varepsilon}^{(m)} \left(1 - r_{\varepsilon}^{(m)}\right)}.
\end{eqnarray}


\vspace{2mm}


\noindent \underline{$X_t(\bm{s}_i), i=1,\ldots, N, t = 1,\ldots,T | rest $} \\
Let the $i$-th entry of $\bm{A} \bm{\varepsilon}^*_t$ be $\tilde{\varepsilon}^*_t(\bm{s}_i)$. The full conditional distribution of $X_t(\bm{s}_i)$ is 
\begin{eqnarray}
\nonumber X_t(\bm{s}_i) | Z_t(\bm{s}_i) = 0, rest &\sim& \textrm{Truncated-Normal}_{(-\infty, 0)}\left(\mu_Z(\bm{s}_i) +  \tilde{\varepsilon}_t(\bm{s}_i), (1 - r_{\varepsilon}) \right), \\
\nonumber X_t(\bm{s}_i) | Z_t(\bm{s}_i) = 1, rest &\sim& \textrm{Truncated-Normal}_{(0, \infty)}\left(\mu_Z(\bm{s}_i) +  \tilde{\varepsilon}_t(\bm{s}_i), (1 - r_{\varepsilon}) \right), \\
\nonumber X_t(\bm{s}_i) | Z_t(\bm{s}_i) = \texttt{NA}, rest &\sim& \textrm{Normal}\left(\mu_Z(\bm{s}_i) +  \tilde{\varepsilon}_t(\bm{s}_i), (1 - r_{\varepsilon}) \right).
\end{eqnarray}

\subsection{Stage 2: Bivariate GMRF modeling for positive CNT and BA data}
We model the positive CNT and BA data jointly using a bivariate model, where we model log-CNT and log-BA as 
\begin{align}
\label{eq:bivarmodel_supp}
 \nonumber \log\{\text{CNT}_t(\bm{s}_i)\} =& \mu_{1}(\bm{s}_i) + \sigma_1(\bm{s}_i) W_{t1}(\bm{s}_i)\\
\log\{\text{BA}_t(\bm{s}_i)\} =& \mu_{2}(\bm{s}_i) + \sigma_2(\bm{s}_i) W_{t2}(\bm{s}_i),
\end{align}
where  $\nonumber \bm{W}_{t}(\bm{s}_i) = [W_{t1}(\bm{s}_i), W_{t2}(\bm{s}_i)]'$ is a bivariate standard Gaussian process with separable spatial covariance structure and independent across time. We approximate the underlying Gaussian process by a GMRF, detailed in the main paper. We draw approximate Bayesian inference using a two-stage method, where first we obtain smoothed estimates of $\mu_{1}(\bm{s}_i)$, $\sigma_1(\bm{s}_i)$, $\mu_{2}(\bm{s}_i)$, and $\sigma_2(\bm{s}_i)$ at each $\bm{s}_i$, using fixed rank kriging.

In the second step, we model the standardized variables $W_{t1}(\bm{s}_i)$ and $W_{t2}(\bm{s}_i)$. We use the same SPDE mesh and the same projection matrix $\bm{A}$ as in Stage 1. Let $\bm{W}_{tp} = [W_{tp}(\bm{s}_i), \ldots, W_{tp}(\bm{s}_N)]'$ for $p=1,2$, and $\bm{W}_{t} = [\bm{W}'_{t1}, \bm{W}'_{t2}]'$. We model $\bm{W}_{t}$ as $\bm{W}_{t} = \left[ \bm{I}_2 \otimes \bm{A} \right] \bm{\eta}^*_{t} + \tilde{\bm{\eta}}_{t}$, where
\begin{eqnarray}
	\nonumber \bm{\eta}^*_{t} \sim \textrm{Normal}_{2N^*}\left(\bm{0}, r_{\eta} \left(
	\begin{array}{c}
	1 ~~ \rho_{\eta} \\
	\rho_{\eta} ~~ 1
	\end{array}
	\right) \otimes \bm{Q}^{-1}_{\phi_{\eta}} \right), ~\textrm{and}~ 
	\end{eqnarray}
	\begin{eqnarray}
	\tilde{\bm{\eta}}_{t} \sim \textrm{Normal}_{2N}\left( \bm{0}, (1 - r_{\eta}) \left(
	\begin{array}{c}
	1 ~~ \rho_{\eta} \\
	\rho_{\eta} ~~ 1
	\end{array}
	\right) \otimes \bm{I}_{N} \right).
	\end{eqnarray}

The set of latent process, parameters, and hyper-parameters in the model are
\begin{eqnarray}
\nonumber && \bm{\Theta}_2 = \left\lbrace \lbrace \bm{\eta}^*_t \rbrace_{t=1}^{T}, \phi_{\eta}, r_{\eta}, \rho_{\eta} \right\rbrace.
\end{eqnarray}

The MCMC steps for updating the parameters in $\bm{\Theta}_2$ and simulating posterior samples from missing $W_{tp}(\bm{s}_i)$ are as follows. For a specific a parameter (or a set of parameters), we call ``$rest$'' the observed data, all the parameters and hyper-parameters in $\bm{\Theta}_2$ except that parameter (or that set of parameters).

Throughout the algorithm, we update the parameters and hyper-parameters consecutively following the full conditional posterior distributions at each iteration as follows:

\noindent \underline{\textbf{$\lbrace \bm{\eta}^*_t \rbrace_{t=1}^{T} | rest$}} \\
The full conditional posterior distribution of $\bm{\eta}^*_t$ is $\bm{\eta}^*_t | rest \sim \textrm{Normal}_{2N^*}(\bm{\mu}^\ast_{\eta^*}, \bm{\Sigma}^\ast_{\eta^*})$, where
\begin{eqnarray}
	\bm{\mu}^\ast_{\eta^*} = \left( \bm{I}_2 \otimes \left[(1 - r_{\eta})^{-1} \bm{A}'\bm{A} + r_{\eta}^{-1} \bm{Q}_{\phi_{\eta}} \right]^{-1} \right) (1 - r_{\varepsilon})^{-1} [\bm{I}_2 \otimes \bm{A}'] \bm{W}_t,
\end{eqnarray}
\begin{eqnarray}
\nonumber && \bm{\Sigma}^\ast_{\eta^*} = \left(
	\begin{array}{c}
	1 ~~ \rho_{\eta} \\
	\rho_{\eta} ~~ 1
	\end{array}
	\right) \otimes \left[(1 - r_{\eta})^{-1} \bm{A}'\bm{A} + r_{\eta}^{-1} \bm{Q}_{\phi_{\eta}} \right]^{-1}.
\end{eqnarray}

The vectors $\bm{W}_t, t = 1, \ldots, T$ are assumed to be independent across $t=1,\ldots, T$, and hence, the full conditional posterior distributions of $\bm{\eta}^*_t; t=1,\ldots, T$ are independent. Thus, $\bm{\eta}^*_t; t=1,\ldots, T$ are updated within MCMC in parallel.

\vspace{2mm}


\noindent \underline{$\phi_{\eta} | rest$} \\
Similar to $\phi_{\varepsilon}$, we consider the prior $\phi_{\eta} \sim \textrm{Uniform}(0, 2\Delta_{\mathcal{S}})$. Let the $m$-th MCMC sample from $\phi_{\eta}$ be denoted by $\phi_{\eta}^{(m)}$. We simulate a candidate sample $\phi_{\eta}^{(c)}$ from $\phi_{\eta}^{(m)}$ following a procedure similar to simulating $\phi_{\varepsilon}^{(c)}$ from $\phi_{\varepsilon}^{(m)}$. The Metropolis-Hastings acceptance ratio is
\begin{eqnarray}
\nonumber \mathcal{R} &=& \ffrac{ \prod_{t=1}^{T} f_{\textrm{Normal}_{2N^*}}\left(\bm{\eta}^*_t; \bm{0}, r_{\eta} \left(
	\begin{array}{c}
	1 ~~ \rho_{\eta} \\
	\rho_{\eta} ~~ 1
	\end{array}
	\right) \otimes \bm{Q}^{-1}_{\phi_{\eta}^{(c)}} \right)}{ \prod_{t = 1}^{T}  f_{\textrm{Normal}_{2N^*}}\left(\bm{\eta}^*_t; \bm{0}, r_{\eta} \left(
	\begin{array}{c}
	1 ~~ \rho_{\eta} \\
	\rho_{\eta} ~~ 1
	\end{array}
	\right) \otimes \bm{Q}^{-1}_{\phi_{\eta}^{(m)}} \right)} \times \ffrac{\phi_{\eta}^{(c)} \left(2\Delta_{\mathcal{S}} - \phi_{\eta}^{(c)}\right)}{\phi_{\eta}^{(m)} \left(2\Delta_{\mathcal{S}} - \phi_{\eta}^{(m)}\right)}.
\end{eqnarray}

\vspace{2mm}


\noindent \underline{$r_{\eta} | rest$} \\
We consider the prior $r_{\eta} \sim \textrm{Uniform}(0, 1)$. Suppose $r_{\eta}^{(m)}$ denotes the $m$-th MCMC sample from $r_{\eta}$. We simulate a candidate sample $r_{\eta}^{(c)}$ from $r_{\eta}^{(m)}$ following a procedure similar to simulating $\phi_{\varepsilon}^{(c)}$ from $\phi_{\varepsilon}^{(m)}$. The Metropolis-Hastings acceptance ratio is 
\begin{eqnarray}
\nonumber \mathcal{R} = \ffrac{\prod_{t=1}^{T} f_{\textrm{Normal}_{2N}}\left(\bm{W}_t; \left[ \bm{I}_2 \otimes \bm{A} \right] \bm{\eta}^*_{t}, (1 - r_{\eta}^{(c)}) \left(
	\begin{array}{c}
	1 ~~ \rho_{\eta} \\
	\rho_{\eta} ~~ 1
	\end{array}
	\right) \otimes \bm{I}_{N} \right)}{ \prod_{t = 1}^{T} f_{\textrm{Normal}_{2N}}\left(\bm{W}_t; \left[ \bm{I}_2 \otimes \bm{A} \right] \bm{\eta}^*_{t}, (1 - r_{\eta}^{(m)}) \left(
	\begin{array}{c}
\nonumber 	1 ~~ \rho_{\eta} \\
	\rho_{\eta} ~~ 1
	\end{array}
	\right) \otimes \bm{I}_{N} \right)} \times \\ \ffrac{ \prod_{t=1}^{T} f_{\textrm{Normal}_{2N^*}}\left(\bm{\eta}^*_t; \bm{0}, r_{\eta}^{(c)} \left(
	\begin{array}{c}
\nonumber 	1 ~~ \rho_{\eta} \\
	\rho_{\eta} ~~ 1
	\end{array}
	\right) \otimes \bm{Q}^{-1}_{\phi_{\eta}} \right)}{ \prod_{t = 1}^{T}  f_{\textrm{Normal}_{2N^*}}\left(\bm{\eta}^*_t; \bm{0}, r_{\eta}^{(m)} \left(
	\begin{array}{c}
\nonumber 	1 ~~ \rho_{\eta} \\
	\rho_{\eta} ~~ 1
	\end{array}
	\right) \otimes \bm{Q}^{-1}_{\phi_{\eta}} \right)} \times \ffrac{r_{\eta}^{(c)}  \left(1 - r_{\eta}^{(c)}\right)}{r_{\eta}^{(m)} \left(1 - r_{\eta}^{(m)}\right)}.
\end{eqnarray}

\vspace{2mm}

\noindent \underline{$\rho_{\eta} | rest$} \\
We consider the prior $\rho_{\eta} \sim \textrm{Uniform}(0, 1)$. Suppose $\rho_{\eta}^{(m)}$ denotes the $m$-th MCMC sample from $\rho_{\eta}$. We simulate a candidate sample $\rho_{\eta}^{(c)}$ from $\rho_{\eta}^{(m)}$ following a procedure similar to simulating $\phi_{\varepsilon}^{(c)}$ from $\phi_{\varepsilon}^{(m)}$. The Metropolis-Hastings acceptance ratio is 
\begin{eqnarray}
\nonumber \mathcal{R} = \ffrac{\prod_{t=1}^{T} f_{\textrm{Normal}_{2N}}\left(\bm{W}_t; \left[ \bm{I}_2 \otimes \bm{A} \right] \bm{\eta}^*_{t}, (1 - r_{\eta}) \left(
	\begin{array}{c}
	1 ~~ \rho_{\eta}^{(c)} \\
	\rho_{\eta}^{(c)} ~~ 1
	\end{array}
	\right) \otimes \bm{I}_{N} \right)}{ \prod_{t = 1}^{T} f_{\textrm{Normal}_{2N}}\left(\bm{W}_t; \left[ \bm{I}_2 \otimes \bm{A} \right] \bm{\eta}^*_{t}, (1 - r_{\eta}) \left(
	\begin{array}{c}
	1 ~~ \rho_{\eta}^{(m)} \\
	\rho_{\eta}^{(m)} ~~ 1
	\end{array}
	\right) \otimes \bm{I}_{N} \right)} \times \\ \ffrac{ \prod_{t=1}^{T} f_{\textrm{Normal}_{2N^*}}\left(\bm{\eta}^*_t; \bm{0}, r_{\eta} \left(
	\begin{array}{c}
	1 ~~ \rho_{\eta}^{(c)} \\
	\rho_{\eta}^{(c)} ~~ 1
	\end{array}
	\right) \otimes \bm{Q}^{-1}_{\phi_{\eta}} \right)}{ \prod_{t = 1}^{T}  f_{\textrm{Normal}_{2N^*}}\left(\bm{\eta}^*_t; \bm{0}, r_{\eta} \left(
	\begin{array}{c}
\nonumber 1 ~~ \rho_{\eta}^{(m)} \\
	\rho_{\eta}^{(m)} ~~ 1
	\end{array}
	\right) \otimes \bm{Q}^{-1}_{\phi_{\eta}} \right)} \times \ffrac{\rho_{\eta}^{(c)}  \left(1 - \rho_{\eta}^{(c)}\right)}{\rho_{\eta}^{(m)} \left(1 - \rho_{\eta}^{(m)}\right)}.
\end{eqnarray}

Further, we also need to simulate missing $W_{tp}(\bm{s}_i)$ from their full conditional posterior distributions. 
Let the elements of $2N$-length vector $\left[ \bm{I}_2 \otimes \bm{A} \right] \bm{\eta}^*_t$ be denoted by $[\bm{\tilde{\eta}}^{*'}_{t1}, \bm{\tilde{\eta}}^{*'}_{t2}]'$, where $\bm{\tilde{\eta}}^{*}_{t1} = [\tilde{\eta}^*_{t1}(\bm{s}_1), \ldots, \tilde{\eta}^*_{t1}(\bm{s}_N)]'$ and $\bm{\tilde{\eta}}^{*}_{t2} = [\tilde{\eta}^*_{t2}(\bm{s}_1), \ldots, \tilde{\eta}^*_{t2}(\bm{s}_N)]'$. The full conditional distribution of $\bm{W}_t(\bm{s}_i) = [W_{t1}(\bm{s}_i), W_{t2}(\bm{s}_i)]'$ is 
\begin{eqnarray}
	\bm{W}_t(\bm{s}_i) \sim \textrm{Normal}_{2}\left( \left(
	\begin{array}{c}
	\tilde{\eta}^*_{t1}(\bm{s}_i)  \\
	 \tilde{\eta}^*_{t2}(\bm{s}_i)
	\end{array}
	\right), (1 - r_{\eta}) \left(
	\begin{array}{c}
\nonumber 1 ~~ \rho_{\eta} \\
	\rho_{\eta} ~~ 1
	\end{array}
	\right) \right).
	\end{eqnarray}

\vspace{2mm}

   
        
\end{document}

%% file: macros.tex
\def\frac#1#2{{\textstyle{#1\over#2}}}

\DeclareSymbolFont{AMSb}{U}{msb}{m}{n}
\DeclareMathSymbol{\Natural}{\mathbin}{AMSb}{"4E}
\DeclareMathSymbol{\Integer}{\mathbin}{AMSb}{"5A}
\DeclareMathSymbol{\Real}{\mathbin}{AMSb}{"52}
\DeclareMathSymbol{\Rational}{\mathbin}{AMSb}{"51}
\DeclareMathSymbol{\Imaginary}{\mathbin}{AMSb}{"49}
\DeclareMathSymbol{\Complex}{\mathbin}{AMSb}{"43} 
\DeclareMathSymbol{\Disk}{\mathbin}{AMSb}{"44} 
\def\bi{\begin{itemize}}
\def\ei{\end{itemize}}
\def\bd{\begin{description}}
\def\ed{\end{description}}
\def\ben{\begin{enumerate}}
\def\een{\end{enumerate}}
\def\bv{\begin{verbatim}}
\def\ev{\end{verbatim}}

\def\hat#1{{\widehat{#1}}}

\def\2to{{\ {\buildrel 2\over \longrightarrow}\ }}

\def\I1ton{{$I_1,\ldots,I_n$}}
\def\X1ton{{$X_1,\ldots,X_n$}}
\def\Y1ton{{$Y_1,\ldots,Y_n$}}
\def\Z1ton{{$Z_1,\ldots,Z_n$}}
\def\R1ton{{$R_1,\ldots,R_n$}}
\def\e1ton{{$e_1,\ldots,e_n$}}
\def\t1ton{{$t_1,\ldots,t_n$}}
\def\x1ton{{$x_1,\ldots,x_n$}}
\def\y1ton{{$y_1,\ldots,y_n$}}
\def\z1ton{{$z_1,\ldots,z_n$}}